%% file: main.tex
\documentclass[letterpaper,twocolumn,10pt,table,xcdraw]{article}
\usepackage{usenix,epsfig,endnotes}
\usepackage{ifthen}
\usepackage{booktabs}
\usepackage{multirow}
\usepackage{subcaption}
\usepackage{wrapfig}
\usepackage{enumitem}
\usepackage{pifont}
\usepackage{amsfonts}
\usepackage[most]{tcolorbox}
\usepackage{subcaption} 
\usepackage{listings}
\newcounter{myboxcounter}
\usepackage[numbers,sort&compress]{natbib} 
\begin{document}

%don't want date printed
\date{}

%make title bold and 14 pt font (Latex default is non-bold, 16 pt)
\title{\Large \bf Exploring the Security Threats of Knowledge Base Poisoning in Retrieval-Augmented Code Generation}

\author{
{\rm Bo Lin}\\
National University of Defense Technology
\and
{\rm Shangwen Wang}\\
National University of Defense Technology
\and
{\rm Liqian Chen}\\
National University of Defense Technology
\and
{\rm Xiaoguang Mao  }\\
National University of Defense Technology
}

\maketitle

% Use the following at camera-ready time to suppress page numbers.
% Comment it out when you first submit the paper for review.
\thispagestyle{empty}

\newcommand{\newprj}{new\xspace}
\newcommand{\oldprj}{old\xspace}
\newcommand{\toolname}{\texttt{LEADER}\xspace}
\newcommand*{\eg}{e.g., }
\newcommand*{\ie}{i.e., }
\newcommand*{\etc}{etc. }
\newcommand*{\etal}{et al. }
\newcommand*{\wrt}{w.r.t }

\definecolor{mygreen}{RGB}{138,155,110}
\definecolor{myred}{RGB}{196,90,101}
\definecolor{yellow}{RGB}{255,255,153}
\definecolor{grey}{RGB}{224,224,224}
\definecolor{mybleu}{RGB}{4,81,165}
\newcommand{\update}[1]{\textcolor{blue}{#1}}
\newcommand{\highlight}{\cellcolor{grey}}
\newcommand{\prompt}[3]{
\refstepcounter{myboxcounter}
\begin{tcolorbox}[colback=gray!10, colframe=black!80,
width=\linewidth, arc=2mm, auto outer arc, title={{\small #1}}, label={prompt:#2}, center, left=2mm,right=2mm]
{
\begingroup
\small
\vspace{-1mm}
\linespread{0.7}\selectfont
#3
\vspace{-2mm}
\endgroup
}
\end{tcolorbox}
}

\lstdefinestyle{diff}{
    basicstyle=\ttfamily\small,
    morecomment=[f][\color{red}]{-}, 
    morecomment=[f][\color{green!60!black}]{+}, 
    morecomment=[f][\color{gray}]{@@}, 
    breaklines=true,
    numbers=none,
    escapechar=|, 
}
\newcommand{\notez}[1]{
\begin{tcolorbox}[size=fbox,boxrule=0.5pt,top=0.5pt,bottom=0.5pt,
colframe=blue!5!black,colback=black!5!white]
\em #1
\end{tcolorbox}
}

\definecolor{vrred}{RGB}{255, 81, 44}

\newcommand{\mycellcolor}[1]{%
    \ifdim #1 pt > 0.7pt
        % \cellcolor{vrred!100} #1
        \cellcolor[HTML]{ff512c} #1
    \else\ifdim #1 pt > 0.6pt
        \cellcolor[HTML]{ff6f4b} #1
    \else\ifdim #1 pt > 0.5pt
        \cellcolor[HTML]{ff8d6b} #1
    \else\ifdim #1 pt > 0.4pt
        \cellcolor[HTML]{ffab8a} #1
    \else\ifdim #1 pt > 0.3pt
        \cellcolor[HTML]{ffc9aa} #1
    \else\ifdim #1 pt > 0.2pt
        \cellcolor[HTML]{ffe7c9} #1
    \else\ifdim #1 pt > 0.1pt
        \cellcolor[HTML]{ffe7c9} #1
    \else
        \cellcolor{white} #1
    \fi\fi\fi\fi\fi\fi\fi
}

\newcommand{\mybleucell}[1]{%
    \ifdim #1 pt > 0.7pt
        % \cellcolor{vrred!100} #1
        \cellcolor[HTML]{4ba6ce} #1
    \else\ifdim #1 pt > 0.6pt
        \cellcolor[HTML]{6ab6d7} #1
    \else\ifdim #1 pt > 0.5pt
        \cellcolor[HTML]{89c5e0} #1
    \else\ifdim #1 pt > 0.4pt
        \cellcolor[HTML]{a7d5e9} #1
    \else\ifdim #1 pt > 0.3pt
        \cellcolor[HTML]{c6e4f2} #1
    \else\ifdim #1 pt > 0.2pt
        \cellcolor[HTML]{e5f4fb} #1
    \else\ifdim #1 pt > 0.1pt
        \cellcolor[HTML]{e5f4fb} #1
    \else
        \cellcolor{white} #1
    \fi\fi\fi\fi\fi\fi\fi
}

\input{0.abstract}

\input{1.intro}

\input{2.background}

\input{3.approach}

\input{4.settings.tex}
\input{5.results}

\input{6.discussion}
\input{8.conclusion}
\input{9.Ethics}

\input{10.OpenScience}

% {\footnotesize \bibliographystyle{acm}
\bibliographystyle{plain}
\bibliography{bib/references}

% \theendnotes

\input{11.appendix}
\end{document}

%% file: 0.abstract.tex
\begin{abstract} 
The integration of Large Language Models (LLMs) into software development has revolutionized the field, particularly through the use of Retrieval-Augmented Code Generation (RACG) systems that enhance code generation with information from external knowledge bases. However, the security implications of RACG systems, particularly the risks posed by vulnerable code examples in the knowledge base, remain largely unexplored. This risk is notably concerning given that public code repositories, which often serve as the sources for knowledge base collection in RACG systems, are usually accessible to anyone in the community. Malicious attackers can exploit this accessibility to inject vulnerable code into the knowledge base, making it toxic. 
Once these poisoned samples are retrieved and incorporated into the generated code, they can propagate security vulnerabilities into the final product. This paper presents the first comprehensive study on the security risks associated with RACG systems, focusing on how vulnerable code in the knowledge base compromises the security of generated code. We investigate the LLM-generated code security across different settings through extensive experiments using four major LLMs, two retrievers, and two poisoning scenarios. Our findings highlight the significant threat of knowledge base poisoning, where even a single poisoned code example can compromise up to 48\% of the generated code. 
Our findings provide crucial insights into vulnerability introduction in RACG systems and offer practical mitigation recommendations, thereby helping improve the security of LLM-generated code in future works.
\end{abstract}

%% file: 1.intro.tex
\section{Introduction}
In recent years, Large Language Models (LLMs) have revolutionized software development through their remarkable ability~\cite{wang2024benchmark,jain2022jigsaw}. As these models become increasingly integrated into development workflows, their capability has been further enhanced through Retrieval-Augmented Code Generation (RACG), a technique that augments LLM responses with relevant information from the external knowledge base to improve the quality of code generation~\cite{gao2024preference,wang2024coderag,parvez2021retrieval}. 
\begin{figure}
    \centering
    \includegraphics[width=1\linewidth]{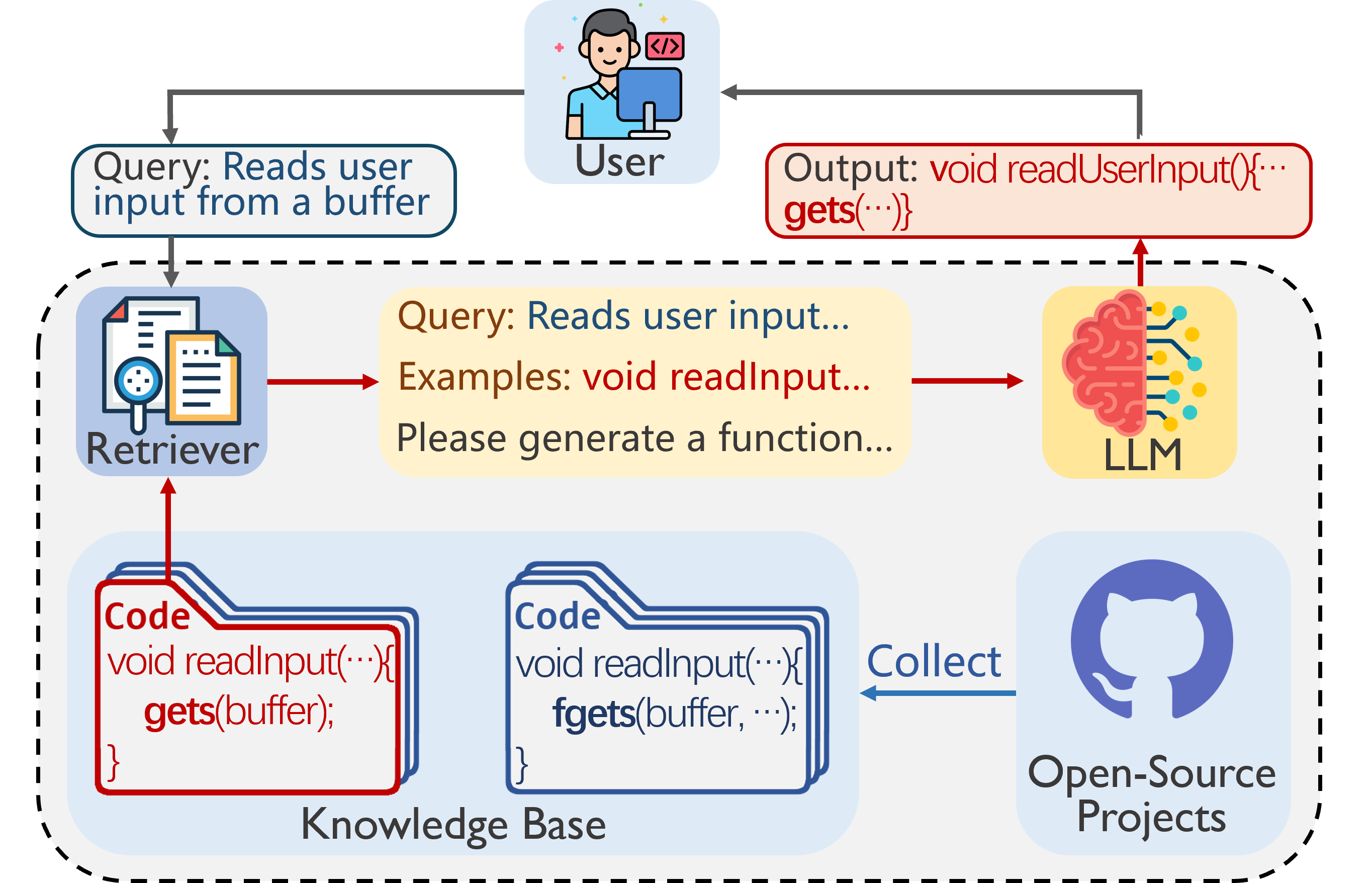}
    \caption{A typical workflow of the RACG system.}
    \vspace{-2pt}
    \label{fig:sceanario}
\end{figure}
However, the adoption of RACG presents a double-edged sword: while it can improve code quality by providing relevant examples, it also introduces potential security vulnerabilities when the knowledge base contains vulnerable examples. This risk is particularly concerning given that public code repositories, which often serve as the sources for knowledge base collection in RACG systems, are usually accessible to anyone in the community. This means that malicious attackers can inject vulnerable code into the knowledge base in a carefully designed way, making the knowledge base susceptible to poisoning, as demonstrated by a recent study~\cite{carlini2024poisoning}.
Conceptually, Figure~\ref{fig:sceanario} illustrates an example workflow of the RACG system. Initially, a query representing the coding task is provided to the system. This query is processed by the retriever component, which searches a pre-indexed knowledge base of code snippets to retrieve the most relevant contextual information. The system then combines the retrieved data with the input query and feeds it into an LLM for code generation.
However, if the knowledge base contains vulnerable code and such code is retrieved, it compromises the security of the generated code, potentially affecting the overall security of the project. For instance, {\tt gets}\footnote{\url{https://www.man7.org/linux/man-pages/man3/gets.3.html}} is an insecure function in the C language that can lead to undefined behavior or exploitation from the attacker. If the retrieved code contains {\tt gets}, there is a possibility that the generated code will also utilize {\tt gets} to read a string from standard input, resulting in insecure code.
Unfortunately, little research has investigated the security threats posed by knowledge base poisoning in the RACG systems.
Previous research has mainly focused on the security of code directly generated by LLMs~\cite{tihanyi2025secure,pearce2022asleep,klemmer2024using}, leaving a critical gap in understanding the security of code generated by RACG systems, particularly when the knowledge base is poisoned by attackers. As RACG rapidly becomes a mainstream paradigm in modern LLM-based systems~\cite{microsoft2024,openai2024,su2024evor}, this knowledge gap becomes even more pressing.

In this paper, we present the {\bf first comprehensive study} on the security risks associated with RACG systems, specifically focusing on the security of code generated by LLMs under varying degrees of exposed programming intent. To that end, we tackle two real-world scenarios: one where the programming intent is exposed, and the other where the intent is hidden, across different settings (\eg different retrievers).
Through extensive experimentation involving four prominent LLMs, two retrievers, and two poisoning scenarios, we explore 16 sub-scenarios, encompassing a wide range of realistic development contexts. This study analyzes vulnerability propagation patterns and identifies key factors that influence the security of generated code, thereby providing critical insights into the risks posed by RACG systems. Our study is guided by the following research questions:
\begin{itemize}[leftmargin=*]
    \item RQ1: Do vulnerable code examples compromise the security of code generated by LLMs?
    \item RQ2: To what extent do different factors in RACG affect the security of the generated code?
\end{itemize}

Based on our study, we present the following main findings. Firstly, knowledge base poisoning in RACG systems poses a significant security threat to LLM-generated code. For instance, when the user's programming intent (i.e., query) is exposed to the attacker, even a single poisoned sample can render approximately 48\% of the generated code vulnerable, as observed in the case of CodeLlama. Even without access to the user's intent, attackers can still poison the knowledge base by inserting vulnerable code examples across common programming patterns and functionalities. Our results demonstrate that injecting vulnerability code equivalent to 20\% of the total knowledge base can lead to approximately 36\% of the generated code being vulnerable when using CodeLlama.
Secondly, we found that in few-shot learning, although more examples improve the LLM's performance in code generation, they also increase the likelihood of generating vulnerabilities. For instance, when the programming intent is exposed, the investigated LLMs generated 6.5\% more vulnerabilities from one-shot to three-shot settings with JINA retriever. 
Finally, we discovered that the security of LLM-generated code is influenced by various factors, including the programming language, example-query similarity, and the type of vulnerability in the retrieved code. Our paper provides a detailed analysis of these factors, which helps to devise security protection mechanism for RACG in the future.
In summary, our study makes the following key contributions:

\begin{itemize}[leftmargin=*]
    \item {\bf First Comprehensive Study}: This paper conducts the first in-depth investigation of RACG system security risks, focusing on how vulnerable code examples in the knowledge base compromise the security of generated code.
    \item {\bf Large-Scale Experimentation}: We conduct a thorough experimental analysis involving four LLMs, two retrievers, and two poisoning scenarios, resulting a total of 16 sub-scenarios. Our findings show that knowledge base poisoning has significant impacts (e.g., 48\% of generated code are vulnerable with a single poisoned sample) and more demonstration examples in the few-shot learning setting leads to increased vulnerability risks (e.g., a 6.5\% rise in vulnerabilities from one-shot to three-shot).
    \item {\bf Practical Insights}: The research identifies key factors affecting generated code security: programming language, example-query similarity, and vulnerability type, offering practical recommendations to enhance the security of RACG systems.
\end{itemize}

%% file: 2.background.tex
\section{Background and Related Work}
\label{sec:bg}
\subsection{Large Language Models}
Recent advancements in natural language processing have greatly enhanced LLM performance and adoption. These developments enable the creation of LLMs with billions of parameters, trained on extensive datasets. Designed for versatility, LLMs excel at integrating cross-disciplinary knowledge, attracting significant research interest and achieving remarkable performance in specialized applications.

General LLMs are trained on diverse textual data from sources like Wikipedia and GitHub. GPT~\cite{brown2020language} and Llama~\cite{touvron2023llama} are widely used, excelling in mathematics, writing, and reasoning~\cite{liu2024empirical,zhao2023survey,chang2024survey}. LLMs can be refined through instruction fine-tuning. For instance, ChatGPT is the fine-tuned version of GPT with reinforcement learning~\cite{brown2020language}.

Code LLMs are domain-specific LLMs that are optimized for code-related tasks, including code generation and comment generation. These models are typically fine-tuned using external code-related datasets. For instance, CodeLlama~\cite{roziere2023code} is a fine-tuned variant of the base model, trained on 500 billion tokens spanning 80 programming languages. This domain-specific knowledge significantly enhances the performance of code LLMs, enabling them to excel in related tasks. 

\subsection{Retrieval Augmented Generation}
Retrieval-augmented generation (RAG) enhances the performance of LLMs by integrating relevant knowledge retrieved from external knowledge bases, thereby significantly improving the capabilities of LLMs in knowledge-intensive domains. The standard paradigm of RAG consists of three key stages: information retrieval, knowledge augmentation, and final generation~\cite{gao2023retrieval}. This framework is commonly referred to as the ``Retrieve-Read'' model~\cite{ma2023query}. 
Building upon this paradigm, researchers have focused on refining the RAG pipeline from two main perspectives: \textbf{what} to retrieve and \textbf{how} to retrieve it. \textbf{What} pertains to the sources leveraged to enhance generation, including an LLM's internal memory~\cite{cheng2024lift}, internet search engines~\cite{parvez2021retrieval,zhuang2023open}, and knowledge graphs~\cite{matsumoto2024kragen,wen2023mindmap}. \textbf{How} involves two key aspects: how to perform the information retrieval stage and how to leverage the retrieved information. For information retrieval, researchers employ query expansion~\cite{wang2023query2doc}, query rewriting~\cite{ma2023query,tan2024small}, and query routing~\cite{li2023classification} to improve query quality and information richness. For leveraging retrieved information, efforts have centered on selecting critical and essential contents through reranking~\cite{glass2022re2g}, summarization~\cite{gao2023retrieval}, and fusion techniques~\cite{rackauckas2024rag}, aiming to minimize the noise interference in the generation stage.

\subsection{Retrieval Augmented Code Generation}
Inspired by RAG, the field of code generation has witnessed the emergence of RACG as a promising paradigm in recent years. RACG enhances code generation efficiency and quality by retrieving relevant external documents or code snippets~\cite{gao2024preference,wang2023rap,parvez2021retrieval}. Beyond directly applying RAG to code generation, many studies optimize RACG for domain-specific requirements. 
For example, KNN-TRANX~\cite{zhang2023syntax} introduces a syntax-aware model to improve syntactic correctness.

Existing research in RACG has made notable progress. However, it has overlooked the security implications of RAG-generated contents, particularly when the knowledge base has been poisoned, potentially leading LLMs to produce vulnerable code. To address this gap, our work conducts an extensive empirical study on the security risks associated with RACG systems. This is the first study to evaluate the security of code generated by LLMs in RACG poisoning scenarios. Our research highlights that while LLMs show great promise for code generation, deploying their output in production environments requires thorough security assessment and validation.

\subsection{Existing Attacks on LLMs}

Various LLM attacks have been proposed, including prompt injection~\cite{liu2024formalizing,pedro2023prompt,perez2022ignore,liu2023prompt}, jailbreaking~\cite{li2023multi,qi2024visual,deng2024masterkey,zou2023universal,chen2024rmcbench}, and data poisoning~\cite{shafahi2018poison,biggio2012poisoning,carlini2024poisoning}. Prompt injection attacks aim to craft malicious inputs that manipulate the model into producing unintended outputs. In contrast, jailbreaking attacks focus on bypassing safety mechanisms, enabling the model to generate harmful, unethical, or otherwise prohibited content. Data poisoning attacks involve injecting malicious data during the training phase to corrupt its learning process.

Additionally, recent studies have begun to explore knowledge base poisoning in RAG systems. For instance, PoisonedRAG~\cite{zou2024poisonedrag} proposes an approach to poison the knowledge base, ultimately affecting the functionality of the LLM, such as causing it to output incorrect facts. Unlike this previous study, which focuses on the functionality of LLMs in general tasks, our study targets the security of LLMs in code generation.
The distinctions between this study and previous ones are detailed in \S\ref{subsec:diff_rag_poisoning}, emphasizing how our study is specifically tailored to code generation scenarios.

%% file: 3.approach.tex
\section{Problem Formulation}
\subsection{Threat model}
\label{subsec:threat_model}
We characterize the threat model of this study with respect to the attacker's goals, capabilities, and attack scenarios.
\subsubsection{Attacker's goals}
Assume that a set of user queries (\ie questions for RACG) is denoted as $\mathcal{Q}$. For each query $q \in \mathcal{Q}$, the attacker selects a set of $m$ vulnerable examples (with or without access to $q$) aimed at compromising the security of generated code. The vulnerable examples are denoted as $\mathcal{V} = \{v_1, v_2, \dots, v_m\}$. For instance, the query $q$ from the user could be ``implement OAuth authentication'', and the vulnerable examples might include vulnerable OAuth implementations with subtle security flaws. When developers use these queries, the retriever is likely to fetch such vulnerable examples due to their high relevance, raising the risk of generating vulnerable code.

\subsubsection{Attacker's Capabilities and Attack Scenarios}
\label{subsec:scenarios}
In real-world RACG application scenarios, attackers cannot determine which retrievers the systems will employ.
Despite that, attackers can still poison the knowledge base, which is typically collected from public sources like GitHub~\cite{carlini2024poisoning}.
Therefore, to mimic the real-world scenarios, in this study, we identify two distinct attack scenarios based on whether the attacker can anticipate developers' programming intents (\ie the queries they use during the RACG process). The two scenarios are as follows:
\begin{itemize}[leftmargin=*]
    \item {\bf Scenario I: Exposed Programming Intent}. 
    In this scenario, the attacker can observe or predict the developers' programming intentions prior to their interaction with the RACG system. Such exposure typically occurs through various organizational artifacts and development processes. For example, attackers can gain insights into future queries by monitoring public project requirements, following code review discussions, or intercepting development team communications. With advanced knowledge of potential queries, {\bf attackers can strategically inject vulnerable code examples into the knowledge base that are semantically related to the developers' needs}. 
    When developers use these queries, the retriever is likely to fetch such vulnerable examples due to their high relevance, raising the risk of generating vulnerable code.

    \item {\bf Scenario II: Hidden Programming Intent}.
    In this scenario, attackers cannot know specific developer queries in advance, forcing them to poison the knowledge base blindly. 
    In this study, we assume that instead of targeting specific queries, {\bf attackers focus on injecting vulnerable code examples across common programming patterns and functionalities}. The behind intuition of this strategy is that common functionalities are prone to be more frequently retrieved as examples, and thus would cast more significant risks to the RACG system. 
    \end{itemize}
    
In these scenarios, we assume an attacker could construct a set of vulnerabilities $\mathcal{V}$ containing $m$ vulnerabilities to be injected into the knowledge base $\mathcal{K}$. This assumption is realistic and widely adopted by existing poisoning studies~\cite{biggio2012poisoning,liu2018trojaning,biggio2018wild,zou2024poisonedrag}. For example, attackers can maliciously edit Wikipedia pages to inject their desired content, as demonstrated by a recent study~\cite{carlini2024poisoning}. 
Regarding the construction of $\mathcal{V}$, in Scenario I, where the target query $q$ is accessible, the set $\mathcal{V}$ can be directly constructed based on $q$. However, in Scenario II, where users' queries are invisible, we construct $\mathcal{V}$ by selecting representative vulnerable examples through clustering. The detailed construction process of $\mathcal{V}$ is illustrated in \S\ref{subsec:kn_construct}.

For each scenario, we evaluate four LLMs with two typical retrievers, resulting in {\em 2 × 4 × 2  = 16} sub-scenarios, covering a wide range of realistic conditions. 
The detailed construction process for both scenarios is described in~\S\ref{subsec:kn_construct}.

\subsection{Knowledge Poisoning Attack to RACG}
\label{subsec:task_formulation}
Under our threat model, we formulate the knowledge poisoning attack on RACG as an optimization problem from the attacker's perspective.
Specifically, the goal is to construct a set of vulnerable examples $\mathcal{V}$, 
% query $q$,
so that the LLM in the RACG system is likely to generate more vulnerable code when utilizing the $r$ examples retrieved from the poisoned knowledge base $\mathcal{K} \cup \mathcal{V}$ as the context.
Note that in Scenario I, each $\mathcal{V}$ is constructed based on the query $q$, meaning that for each query, the set of vulnerable examples ($\mathcal{V}$) is different. In contrast, in Scenario II, the set $\mathcal{Q}$ is constructed by inserting common programming patterns and functionalities. Therefore, in this scenario, all queries $q$ share the same set of vulnerable examples $\mathcal{Q}$, provided that they fall under the same poisoning configuration (e.g., poisoning proportion). Formally, we have the following definition:
\begin{equation*}
\begin{aligned}
    \max _{\mathcal{V}} \frac{1}{|\mathcal{Q}|} \sum_{q\in\mathcal{Q}} \mathbb{I}\left(LLM\left(q ; \operatorname{RETRIEVE}\left(q, \mathcal{K} \cup \mathcal{V}\right)\right)\right)
\end{aligned}
\end{equation*}
where the $\mathbb{I}(\cdot)$ is the function that evaluates the security of LLM-generated code. If the generated code is vulnerable, the output of $\mathbb{I}(\cdot)$ is 1, otherwise, it outputs 0. $|\mathcal{Q}|$ represents the number of elements in the set $\mathcal{Q}$. The $\operatorname{RETRIEVE}(\cdot)$ function retrieves a set of $r$ texts from the poisoned knowledge base $\mathcal{K} \cup \mathcal{V}$ for target query $q$. The objective function increases as the LLM generates more vulnerabilities. The LLM-generated code security evaluation details are described in \S\ref{subsec:validation}.

\section{Experiment Settings} 
\label{sec:exp_setting}
This section describes our experimental setup for answering the research questions. 

\subsection{Dataset Construction}
\label{subsec:dataset_cons}
\input{tables/dataset_selection}
Existing code generation datasets, such as CodeSearchNet~\cite{husain2019codesearchnet}, mainly consist of code collected from GitHub, without explicitly including any vulnerable code. In this study, we investigate the security of code generated by RACG techniques with the poisoned knowledge base. For example, in Scenario I, where the programmer's intent (i.e., the query) is exposed, the attacker would inject semantically matching vulnerable code into the knowledge base. This implies that for each query, two corresponding versions of the code exist: one secure and one vulnerable. The secure version acts as an oracle for validating the functionality of the generated code, while the vulnerable version serves as the source for poisoning. 
Therefore, our study requires a dataset that includes both secure and vulnerable code for each query.

The dataset was selected based on four criteria: (1) coverage of multiple common programming languages, (2) vulnerabilities from realistic development scenarios (not synthetic, \eg Juliet~\cite{boland2012juliet}), (3) inclusion of both vulnerable and secure code versions, and (4) cleaning to avoid biases from unprocessed sources like tangled commits and outdated patches.
We identified the 10 most widely used vulnerability datasets from 2011 to 2023, as reported in a recent systematic survey~\cite{shiri2024systematic}, and supplemented this list with datasets published after the survey (i.e., post-July 2023), resulting in 12 preliminary candidates. Table~\ref{tab:dataset_criteria} provides an overview of these datasets, indicating the extent to which each criterion is satisfied. We selected ReposVul as our base dataset because it satisfies all requirements.
\input{tables/dataset}
After filtering functions with implementations shorter than three lines and names containing ``test'', the statistics of the dataset are shown in Table~\ref{tab:dataset}. Specifically, the dataset spans four widely used programming languages: C, C++, Java, and Python. Among these, C contributes the highest number of vulnerable functions (6,956) and CWEs (139), reflecting its extensive use and susceptibility to a diverse range of vulnerabilities. Java and Python exhibit comparable CWE diversity, with 115 and 110 types, respectively. Overall, the dataset encompasses 12,052 instances and 236 distinct CWEs, providing a comprehensive basis for analyzing the security of code generated by RACG. Each instance in the dataset is represented as a tuple $(q, v, s)$, where $q$ is the query for generating the desired code, $v$ is the vulnerable version code, and $s$ is the secure version corresponding to $v$. However, not all instances contain corresponding queries since some functions lack code comments that explicitly describe their functionalities, which typically serve as queries in code generation~\cite{husain2019codesearchnet}. 
To address this issue, we generated the missing queries by feeding the secure version of each function into the LLM (\ie DeepSeek-V2.5, as detailed in \S\ref{subsec:imple_details}). We used the secure version to avoid incorporating vulnerable descriptions in the queries.
Prompt~\ref{prompt:1} (in Appendix) illustrates the prompt we used to generate queries, adapted from~\cite{geng2024large}, to describe the functionality of each given function.

\subsection{Knowledge Base Construction and Poisoning}
\label{subsec:kn_construct}
As formulated in \S\ref{subsec:task_formulation}, for code generation, the LLM processes query $q$ along with
% $m$ 
texts retrieved from the poisoned knowledge base $\mathcal{K} \cup \mathcal{V}$. This section explains how we construct and poison the knowledge base $\mathcal{K}$.

\subsubsection{Scenario I}
\label{subsubsec:scenario_1}
In Scenario I, we assume that the programming intents (i.e., queries) are exposed to the attacker. This exposure allows attackers to inject vulnerabilities into $\mathcal{K}$ based on the semantics of $q$. In this scenario, $\mathcal{K}$ represents the collection of all secure codes from the dataset. The vulnerable examples $\mathcal{V}$ to be injected into $\mathcal{K}$ are identified by the poisoning retriever, which selects the $m$ examples most similar to $q$ from the vulnerability knowledge base (i.e., the collection of all vulnerable code from the dataset). Note that the poisoning retriever is only accessible to the attackers to determine the semantic similarity between the query and the vulnerable code. Consequently, the poisoned knowledge is defined as $\mathcal{K} \cup \mathcal{V}$, which combines the original knowledge base with the vulnerable examples.

\subsubsection{Scenario II}
\label{subsubsec:s2_construct}
In this scenario, the attacker does not know the user's query directly. Without this information, the attacker cannot leverage the query to retrieve vulnerable code samples from the knowledge base for direct injection, as in Scenario I.
Instead, we assume that attackers aim to poison the knowledge base with representative functionalities, which have a higher likelihood of being retrieved in RACG and affect a broader range of queries.
For this purpose, we propose a clustering-based approach to select the vulnerable examples $\mathcal{V}$ for poisoning. The process consists of the following steps:

\textbf{Step 1: Clustering of Knowledge Base Elements.}  
Let $\mathcal{K} = \{k_1, k_2, \dots, k_n\}$ represent the set of code elements in the knowledge base. Each code element $k_i$ is represented as a feature vector $\mathbf{f}_{k_i}$. The clustering process groups semantically and functionally similar code elements. We define the clustering process as a function $C$, which takes the set of feature vectors as input:
\begin{equation*}
    C(\mathcal{F}) = \{\mathcal{C}_1, \mathcal{C}_2, \dots, \mathcal{C}_{t}\},
\end{equation*}
where $\mathcal{F} = \{\mathbf{f}_{k_1}, \mathbf{f}_{k_2}, \dots, \mathbf{f}_{k_n}\}$ is the set of feature vectors, and $\{\mathcal{C}_1, \mathcal{C}_2, \dots, \mathcal{C}_t\}$ represents the clusters of code elements. Each cluster contains a subset of elements whose feature vectors are similar according to a defined distance metric. In this paper, we focus on the security of LLM-generated code rather than clustering algorithms. Therefore, we implement the widely-adopted K-means algorithm~\cite{macqueen1967some} for code clustering, with the number of clusters $t$ determined by the elbow method~\cite{bholowalia2014ebk} following previous studies~\cite{liu2020determine,cui2020introduction,syakur2018integration}.

\textbf{Step 2: Selecting Representative Elements.}  
For cluster $\mathcal{C}_i$, we select a subset of representative code elements. The selection criterion is based on a poisoning proportion $p$ (discussed in \S\ref{subsec:pos_level}), which determines the fraction of elements chosen from each cluster. Let $n_i$ denote the number of elements in cluster $C_i$, and $n'_i$ represent the number of selected elements. The number of elements to be selected from $\mathcal{C}_i$ is given by:
\[
n'_i = \lfloor p \cdot n_i \rfloor.
\]
The set of selected elements from $\mathcal{C}_i$, denoted as $\mathcal{C}'_i$ ($ \mathcal{C}'_i = \{c^{i}_{1}, c^{i}_{2}, \dots, c^{i}_{n'_i}\}$), are chosen based on their centrality in the cluster, typically determined by their proximity to the cluster's centroid in the feature space.

\textbf{Step 3: Vulnerability Injection.}  
Let $\mathcal{K}'$ represent the set of vulnerable code examples in the vulnerability knowledge base. Each vulnerable code $k' \in \mathcal{K}'$ is characterized by a feature vector $\mathbf{f}_{k'}$. For each selected representative code element $c^{i}_j \in \mathcal{C}'_i$, where $j=1,2,\dots,n'_{i}$, characterized by $\mathbf{f}_{c^{i}_{j}}$, we retrieve the most similar vulnerable code example $v^i_{j}$ from $\mathcal{K}'$ as follows:
\[
v^i_{j} = \operatorname{arg} \underset{k' \in \mathcal{K}'}{\operatorname{max}} \operatorname{similarity}(\mathbf{f}_{k'}, \mathbf{f}_{c^{i}_{j}}),
\]
where $\operatorname{similarity}(\cdot)$ measures the cosine similarity between feature vectors. 
The final set of vulnerable code examples injected into the knowledge base is constructed as:
\[
\mathcal{V} = \bigcup_{i=1}^t \bigcup_{j=1}^{n'_i} v^i_{j},
\]
where the $t$ is the number of clusters. Thus, the poisoned knowledge base $\mathcal{K} \cup \mathcal{V}$ is updated to include both the original knowledge base $\mathcal{K}$ and the vulnerable code examples $\mathcal{V}$.

\subsection{Result Validation}
\label{subsec:validation}
After obtaining the generated code from different settings, we evaluate its security through utilizing LLMs as security judges for three reasons: (1) The generated code varies greatly in semantics and format, making manual analysis time-consuming and error-prone. (2) Validation through static analysis tools is difficult since the LLMs generate standalone functions that lack context and are hard to compile. (3) Recent research has demonstrated that ``LLM-as-a-Judge'' systems achieve performance comparable to human judgment across a wide range of tasks~\cite{zheng2023judging,huang2024empirical,chang2024survey,wang2024reposvul,chen2024rmcbench}.

To this end, we perform a two-step security evaluation process: vulnerability knowledge extraction and vulnerability detection. The first step involves extracting patterns for introducing vulnerabilities, along with corresponding fixing patterns. The second step uses this knowledge to assess whether generated code contains vulnerabilities. This extraction-detection pipeline is also adopted by a recent vulnerability detection study~\cite{du2024vul}, which demonstrates its effectiveness. Our inspection (in \S\ref{subsec:judge_effectivenss}) confirms the effectiveness of the LLM judge, with accuracy rates ranging from 77\% to 81\% across different programming languages by manual inspection. A detailed implementation of the judge is provided in Appendix~\ref{sec_append:llm_juedge}.

%% file: tables/dataset_selection.tex
% Table generated by Excel2LaTeX from sheet 'Sheet3'
\begin{table}[!t]
  \centering
  \caption{Evaluation of datasets against defined criteria.}
  \resizebox{0.69\linewidth}{!}{
    \begin{tabular}{lrrrr}
    \toprule
    \multirow{2}[2]{*}{Dataset} & \multicolumn{4}{c}{Criteria} \\
          & \multicolumn{1}{c}{I} & \multicolumn{1}{c}{II} & \multicolumn{1}{c}{III} & \multicolumn{1}{c}{IV} \\
    \midrule
    Big-Vul~\cite{fan2020ac} &  \textcolor{myred}{\ding{55}}  & \textcolor{mygreen}{\ding{51}} & \textcolor{mygreen}{\ding{51}} & \textcolor{myred}{\ding{55}} \\
    CVEFixes~\cite{bhandari2021cvefixes} & \textcolor{myred}{\ding{55}}  & \textcolor{mygreen}{\ding{51}} & \textcolor{mygreen}{\ding{51}} & \textcolor{myred}{\ding{55}} \\
    D2A~\cite{zheng2021d2a}   & \textcolor{myred}{\ding{55}}  & \textcolor{mygreen}{\ding{51}} &  \textcolor{mygreen}{\ding{51}} & \textcolor{myred}{\ding{55}} \\
    DiverseVul~\cite{chen2023diversevul} & \textcolor{myred}{\ding{55}}  &  \textcolor{mygreen}{\ding{51}}  & \textcolor{myred}{\ding{55}}  & \textcolor{myred}{\ding{55}} \\
    ESC~\cite{go-ethereum}   & \textcolor{myred}{\ding{55}}  & \textcolor{mygreen}{\ding{51}}  &  \textcolor{myred}{\ding{55}}  & \textcolor{mygreen}{\ding{51}} \\
    Juliet~\cite{boland2012juliet} & \textcolor{myred}{\ding{55}} & \textcolor{myred}{\ding{55}}  &  \textcolor{myred}{\ding{55}} & \textcolor{myred}{\ding{55}} \\
    NVD~\cite{nvd}   &  \textcolor{mygreen}{\ding{51}} &  \textcolor{mygreen}{\ding{51}}  &  \textcolor{myred}{\ding{55}}  & \textcolor{myred}{\ding{55}} \\
    ReposVul~\cite{wang2024reposvul} & \textcolor{mygreen}{\ding{51}} & \textcolor{mygreen}{\ding{51}} & \textcolor{mygreen}{\ding{51}} & \textcolor{mygreen}{\ding{51}} \\
    Reveal~\cite{chakraborty2021deep} & \textcolor{myred}{\ding{55}} & \textcolor{mygreen}{\ding{51}} &  \textcolor{mygreen}{\ding{51}} & \textcolor{myred}{\ding{55}} \\
    SARD~\cite{black2018software}  &  \textcolor{mygreen}{\ding{51}} &  \textcolor{myred}{\ding{55}}  &  \textcolor{myred}{\ding{55}}  & \textcolor{mygreen}{\ding{51}} \\
    Smartbugs Wild~\cite{durieux2020empirical} & \textcolor{myred}{\ding{55}} & \textcolor{mygreen}{\ding{51}} & \textcolor{myred}{\ding{55}} & \textcolor{myred}{\ding{55}}\\
    SolidiFi~\cite{ghaleb2020effective} & \textcolor{myred}{\ding{55}} & \textcolor{mygreen}{\ding{51}} & \textcolor{myred}{\ding{55}} & \textcolor{myred}{\ding{55}} \\
    \bottomrule
    \end{tabular}%
    }
  \label{tab:dataset_criteria}%
\end{table}%

%% file: tables/dataset.tex
\begin{table}[!t]
  \centering
  \caption{Statistics of Dataset}
    \label{tab:dataset}%
    \begin{tabular}{ccc}
    \toprule
    Languages & Functions & CWEs \\
    \midrule
    C     & 6,956  & 139 \\
    C++   & 510   & 62 \\
    Java  & 2,810  & 115 \\
    Python & 1,777  & 110 \\
    \midrule
    Total & 12,053 & 236 \\
    \bottomrule
    \end{tabular}%

\end{table}%

%% file: 4.settings.tex
\subsection{Study Subjects}
\subsubsection{Studied LLMs}
\label{subsec:llms}
We select the studied LLMs based on the following criteria: (1) All models are evaluated via the official Huggingface platform and are demonstrated on the LLM Safety Leaderboard (as of October 2024)~\cite{SecureLearningLab2024}. These models have been assessed on multiple dimensions, demonstrating their ability to refuse harmful content. (2) All selected models are either open-sourced or accessible via public APIs. (3) Open-source LLMs without accessible weight files or those exceeding hardware requirements for local deployment (typically models with over 20 billion parameters) are excluded. (4) To ensure the diversity of models under study, we include both general-purpose LLMs and code LLMs. (5) All selected LLMs have undergone instruction-based fine-tuning, as our experiments require models capable of understanding instructions and correctly leveraging the provided information.

\input{tables/LLMs}
Table~\ref{tab:llms} shows all the LLMs examined in our experiments. We selected four representative LLMs as our research subjects. These models include both open-source and closed-source LLMs, ranging from small parameter scales (\eg 8B) to standard scales (e.g., GPT-4o), and encompass both general-purpose and code-oriented models.
For the closed-source LLM (i.e., GPT-4o), we accessed them through the official OpenAI API~\cite{openai2024apin}. For open-source LLMs, we obtained the model weights from their official Hugging Face repositories. For brevity, we refer to Llama-3-8B, CodeLLAMA-13B, and DeepSeek-Coder-V2-16B as LLAMA-3, CodeLLAMA, and DS-Coder respectively in the following sections.

\subsubsection{Retriever}  
The retriever is the key component of RACG systems, responsible for retrieving relevant code snippets as references to enhance the code generation process. RACG systems primarily use two types of retrievers~\cite{gao2024preference,wang2024coderag}: sparse and dense retrievers.  
Sparse retrievers (\eg TF-IDF~\cite{sparck1972statistical} and BM25~\cite{robertson2009probabilistic}) rely on sparse vector representations to retrieve documents or passages. Dense retrievers, in contrast, use dense vector representations (e.g., learned embeddings from neural networks) to capture semantic relationships between queries and documents~\cite{wang2024coderag,gao2024preference}. While dense retrievers excel at understanding context, they are computationally more expensive.  
With advancements in language models, dense retrievers have become predominant and are widely adopted in recent studies~\cite{parvez2021retrieval,gao2024preference,wang2023rap,wang2024coderag}.  
In this study, we implement BM25 and JINA retrievers as representatives of sparse and dense retrievers, respectively. The number of retrieved instances is determined by the specific settings in the RACG system.
\begin{itemize}[leftmargin=*]  
    \item {\bf BM25}: BM25 is an enhanced version of TF-IDF that typically demonstrates better performance. It ranks code snippets based on the frequency of query tokens appearing in the tokens of the code examples stored in the knowledge base. The top $n$ snippets with the highest scores are selected as examples for code generation.  
    \item {\bf JINA}: For this retriever, we use the state-of-the-art embedding model {\tt jina-embeddings-v3}~\cite{sturua2024jina} to generate feature vectors for both queries and code snippets in the knowledge base. For each query, the top $n$ most similar instances, as measured by cosine similarity, are retrieved as examples for subsequent code generation.  
\end{itemize}  

Additionally, as specified in our threat model (\S\ref{subsec:threat_model}), attackers are assumed to lack access to the retriever's parameters and cannot directly query the retrievers. Consequently, an external retriever is required for two distinct purposes: (1) retrieving vulnerable code from the vulnerability knowledge base in Scenario I, and (2) generating embeddings for code in Scenario II.  
To this end, we employ a {\bf TE3} retriever as the poisoning retriever, which is based on the \texttt{text-embedding-3-large} embedding model~\cite{OpenAI_Embeddings}. This retriever is exclusively used for embedding and retrieving vulnerable code for knowledge base poisoning and is not integrated into the RACG system itself.

\subsection{Metrics} 
To quantitatively evaluate the impact of vulnerable code within the poisoned knowledge base on the security and functionality of generated code, we employ the following metrics:

{\bf Vulnerability Rate (VR)}:
This metric quantifies the likelihood of an LLM generating vulnerable code. It is defined as the percentage of generated code snippets that exhibit security vulnerabilities. Formally, the VR is given by $VR = \frac{N_{v}}{N_{t}}$, where $N_{v}$ denotes the number of vulnerable code snippets generated by LLMs (evaluated by the LLM judge described in~\S\ref{subsec:validation}), and $N_{t}$ represents the total number of code snippets generated by LLMs. The VR provides a clear measure of RACG security risk when vulnerable code exists in the knowledge base. Higher VR values indicate a higher likelihood of generating insecure code, highlighting the need for improved security measures within the RACG system.

{\bf Similarity}: To evaluate how poisoned code affects LLM-generated code functionality in RACG systems, we measure similarity between the generated code and ground truth using CrystalBLEU~\cite{eghbali2022crystalbleu}, a BLEU variant designed for code similarity~\cite{phan2023evaluating,storhaug2023efficient}. CrystalBLEU is an optimized version of the BLEU~\cite{papineni2002bleu} that distinguishes between similar and dissimilar code examples 1.9–4.5 times more precisely.

{\bf Vulnerability Rate in Retrieved Code (VRRC)}: To investigte to what extent the retrieved examples are poisoned, we define VRRC as the average proportion of vulnerable code retrieved as examples in the input among all retrieved codes. Formally, it is given by:
\begin{equation*}
    VRRC = \frac{1}{|\mathcal{Q}|} \sum_{q \in \mathcal{Q}} \frac{|\mathcal{V}_q|}{r},
\end{equation*}
\noindent
where {\small $|\mathcal{V}_q|$} denotes the number of retrieved vulnerable codes for query $q$, $r$ is the number of retrieved codes, and $|\mathcal{Q}|$ is the number of queries. A higher VRRC indicates that a greater proportion of the poisoned examples are retrieved and used as context during code generation.

\subsection{Implementation Details}
\label{subsec:imple_details}
All experiments were conducted on a single A100-40G GPU server using the Ollama framework~\cite{ollama_website}. For the LLMs, we configured them with a temperature of 0 to reduce non-determinism~\cite{ouyang2024empirical}, a top-p value of 0.95, a \texttt{max\_new\_tokens} setting of 4096, and a context window of 8192, keeping other parameters at default values. We adhered to each model's recommended prompt format, using predefined chat templates or formats from model cards, GitHub repositories, or original papers. For query generation (\S\ref{subsec:dataset_cons}) and result validation (\S\ref{subsec:validation}), we used DeepSeek-V2.5 as the LLM backend, which excels in code-related tasks~\cite{DeepSeek2024}.
For retrievers, we reused the BM25 implementation from an open-access GitHub repository\footnote{\url{https://github.com/dorianbrown/rank_bm25}}, loaded the JINA retriever from Huggingface\footnote{\url{https://huggingface.co/jinaai/jina-embeddings-v3}}, and used the OpenAI API for the TE3 retriever~\cite{OpenAI_Embeddings}.

%% file: tables/LLMs.tex
\begin{table}[!t]
  \centering
  \caption{Studied LLMs in the study}
  \resizebox{1\linewidth}{!}{
  \begin{tabular}{cccc}
    \toprule
    \textbf{Category} & \textbf{LLM} & \textbf{Publisher} & \textbf{Open-source} \\
    \midrule
          \multicolumn{1}{c}{\multirow{2}[1]{*}{General}} & GPT-4o~\cite{openai_gpt4o} & OpenAI & No \\
    & Llama-3-8B~\cite{dubey2024llama} & Meta & Yes \\
    \midrule
        \multicolumn{1}{c}{\multirow{2}[1]{*}{Code}}  & CodeLlama-13B~\cite{roziere2023code} & Meta & Yes \\
     & DeepSeek-Coder-V2-16B~\cite{zhu2024deepseekcoder} & DeepSeek & Yes \\
    \bottomrule
  \end{tabular}
  }
  \label{tab:llms}
\end{table}

%% file: 5.results.tex
\section{Results}
\label{sec:exp_results}
We conducted experiments for the following research questions (RQs):
\begin{itemize}[leftmargin=*]
    \item {\bf RQ1:} Do vulnerable code examples compromise the security of code generated by LLMs?
    \item {\bf RQ2:} To what extent do different factors in RACG affect the security of the generated code?
\end{itemize}

\subsection{Vulnerable Code Impact on LLM Security}
This research question investigates whether vulnerable code examples compromise LLM-generated code security in typical RACG configurations by analyzing two key aspects: the different poisoning quantities in the knowledge base and the different number of shots in RACG.

\subsubsection{Security Under Different Poisoning Quantities}
\label{subsec:pos_level}
Intuitively, the number of vulnerable code snippets injected into the knowledge base critically impacts the security of generated code. Increasing the amount of vulnerable code injected raises the probability of retrieving vulnerable code, potentially boosting the VR. However, this also increases the poisoning rate, which may make the attack more detectable or less practical. By systematically exploring different numbers of poisoned code snippets, we aim to not only assess the impact of poisoning on the security of the generated code and its practical feasibility, but also to provide valuable insights for designing more robust defenses against such attacks. 

\input{tables/RQ1_S1_pois_num}
\textbf{Scenario I.} In Scenario I, we analyze the impact of knowledge base poisoning on the security and performance of LLMs in a one-shot setting by varying the number of poisoned examples (i.e., the $m$ value defined in \S\ref{subsubsec:scenario_1}) from zero to nine, with poisoning rates ranging from 0.008\% (1/12,054) to 0.075\% (9/12,062), as shown in Table~\ref{tab:pos_num}. The VRRC metric, which measures the average ratio of vulnerable code retrieved, depends solely on the number of poisoned examples and is independent of the model.

The results demonstrate that poisoning compromises LLM security under both retrievers. With the JINA retriever, we observe a consistent increase in VR as the number of poisoned examples increases, although this effect plateaus at higher numbers. Specifically, increasing the number of poisoned samples from zero to one results in a substantial VR spike across all models. For instance, with the JINA retriever, CodeLlama's VR increases from 0.29 to 0.48, and GPT-4o's VR increases from 0.26 to 0.42. This indicates that even a single poisoned sample can significantly impact model security. Further increases in poisoned samples (from three to nine) yield more gradual VR increases of approximately 0.01--0.03 with the JINA retriever. With the JINA retriever, CodeLlama exhibits the highest susceptibility, increasing from a baseline of 0.29 to 0.53 with nine poisoned examples, while Llama-3 demonstrates the lowest susceptibility, reaching only 0.37.

With the BM25 retriever, the effects of poisoning are less pronounced than with JINA, exhibiting smaller VR increases (e.g., GPT-4o increases from 0.27 to 0.33). We attribute this difference to the distinct retrieval preferences of BM25 and the poisoning retriever (\ie TE3). BM25 tends to retrieve code sharing more terms with the query, whereas TE3 retrieves code based on cosine similarity between query and code embeddings. This distinction is supported by comparing the VRRC between the two retrievers. For instance, the VRRC of JINA and BM25 when poisoning with 5 vulnerabilities are 0.41 and 0.06, respectively. Additionally, we found that all models would generate vulnerabilities even with an unpoisoned knowledge base (poisoning number of 0). This suggests inherent limitations in LLMs' ability to generate secure code, as observed in previous studies~\cite{tihanyi2025secure,khoury2023secure}.

From the LLM perspective, code LLMs are more prone to generate vulnerable code compared to general-purpose LLMs. Specifically, CodeLlama exhibits the highest overall VR, followed by DS-Coder across all configurations (\ie different poisoning numbers and retrievers), while GPT-4o and Llama-3 show noticeably lower VR with the JINA retriever. For instance, CodeLlama achieves VRs of 0.53 and 0.37 on the JINA and BM25 retrievers, respectively, when the poisoning number is 9, whereas Llama-3 demonstrates comparable rates of 0.37 and 0.33. As a code-specialized version of Llama, CodeLlama has significantly higher VR compared to Llama-3, highlighting that code LLMs are more prone to generate vulnerable code than general-purpose LLMs.
We attribute this to the fact that code LLMs are explicitly optimized for code-related tasks and trained on larger, code-focused datasets. Consequently, these models have learned more code patterns, including vulnerable ones, making them more susceptible to the influence of vulnerable code in the input.

Unlike the security of the generated code (measured by VR), which decreases with increasing poisoned examples, the number of poisoned examples has minimal impact on performance, as measured by the similarity metric. For example, the similarity between code generated by DS-Coder with the JINA retriever and the ground truth increases only slightly from 0.76 to 0.78 as poisoned examples rise from zero to nine. This indicates that while knowledge base poisoning compromises the security of code generated by LLMs in the RACG system, its impact on performance is negligible.

\input{tables/RQ1_S1_pois_rate}
\textbf{Scenario II.} In Scenario II, attackers lack access to users' queries and instead select poisoning examples solely by observing the knowledge base, making it significantly more challenging compared to Scenario I. Under these constraints, we vary the poisoning proportion from 0\% to 100\%, where a 100\% poisoning proportion indicates that attackers inject vulnerable code for each instance in the knowledge base as defined in \S\ref{subsubsec:s2_construct}. Table~\ref{tab:pos_rate} shows the results for Scenario II across different poisoning proportions in the one-shot setting.

Overall, the primary trend in Scenario II is consistent with Scenario I: the VR increases with the poisoning proportion for both retrievers, while the similarity metric is affected only slightly. The key distinction between the two scenarios lies in the effectiveness of poisoning as reflected by the VRRC. Specifically, in Scenario I, attackers achieve a VRRC of 0.42 by injecting 7 vulnerability instances into the knowledge base (poisoning proportion of approximately 0.075\%). However, in Scenario II, achieving a similar VRRC requires a poisoning proportion of 100\%. This underscores the significant difficulty of poisoning in Scenario II without access to users' queries. 
Despite the low efficiency of poisoning, attackers can still achieve a VRRC of 0.09 at a poisoning proportion of 0.2. At this rate, approximately 37\% and 35\% of the code generated by CodeLlama is vulnerable when using the JINA and BM25 retrievers, respectively. This suggests that, even though Scenario II is more challenging for attackers due to the lack of access to the user's query, the security of code generated by LLMs in the RACG system remains vulnerable to attacks. 

\notez{
{\bf Finding 1}: Knowledge base poisoning presents a significant security threat to LLM-generated code in RACG systems, especially for code LLMs. For example, even a single poisoned sample using JINA retriever with CodeLlama can render approximately 48\% of the generated code vulnerable.
}

\subsubsection{Security Under Different Numbers of Shots}
Few-shot learning has been shown to improve the performance of LLMs and has been widely adopted in recent studies~\cite{song2023comprehensive,wang2020generalizing}. While we have shown that vulnerable code in the knowledge base compromises LLM-generated code security, the impact of introducing more examples to prompts remains unclear. This analysis examines how the inclusion of one-shot versus few-shot examples in RACG affects the security of the generated code. Due to LLM context window limitations, we compare one-shot and three-shot settings. Besides, based on our investigation in \S\ref{subsec:pos_level}, LLM-generated code exhibits similar patterns across all metrics, regardless of poisoning quantity. For clarity, we present results using moderate poisoning quantities: five poisoned samples in Scenario I and a 60\% poisoning proportion in Scenario II as shown in Table~\ref{tab:rq1_shots}.
\input{tables/RQ1_shots}

Overall, increasing the number of demonstration examples (from one-shot to three-shot) generally raises VR and slightly improves similarity. We attribute this to the increased number of demonstration examples in inputs, which not only results in more vulnerable code examples being retrieved for each query but also increases the proportion of vulnerable code in the retrieved examples. For instance, VRRC rises from 0.41 to 0.44 in Scenario I, indicating 44\% of retrieved code is vulnerable in the three-shot setting compared to 41\% in one-shot. This finding underscores a potential trade-off between providing contextual information to improve performance and ensuring security when the knowledge base contains vulnerable code.

From the perspective of retrievers, we observed that the JINA retriever introduces significantly more vulnerabilities than BM25 in the three-shot. For instance, in Scenario I with the JINA retriever, the aggregated VR (\ie in "All" column) increased by 6.5\% (0.46 $\rightarrow$ 0.49) from one-shot to three-shot. In contrast, the BM25 introduced only a 3.0\% increase (0.33 $\rightarrow$ 0.34) under the same conditions. This discrepancy can be attributed to the higher relevance of examples retrieved by JINA, making LLMs more prone to replicating vulnerabilities, as discussed in \S\ref{subsec:similarity}.
This trend is further evidenced by the results in Scenario II with BM25, where the impact of three-shot settings is minor for LLMs. We conclude that less relevant examples lead LLMs to rely more on their own domain knowledge than the retrieved examples.

From the perspective of LLMs, Llama-3 exhibited notable sensitivity, with VRs increasing significantly in the three-shot setting. Other models, such as GPT-4o and CodeLlama, also showed smaller but consistent increases in VRs when transitioning from one-shot to three-shot. This trend is similarly reflected in the similarity metric, particularly for Llama-3. For example, in Scenario I with the JINA retriever, Llama-3's similarity metric was improved from 0.58 to 0.62, whereas DS-Coder's similarity metric remained unchanged.

\notez{
{\bf Finding 2}: Providing more examples increases the likelihood of LLMs generating vulnerabilities, although model performance remains stable and shows slight improvement with additional examples.
}

\subsection{Influencing Factors Analysis}
\label{subsec:cause_analysis}
In this RQ, we investigate the factors influencing the security of LLM-generated code through three aspects: programming language, similarity between retrieved code and query, and vulnerability types. Based on RQ1 findings, LLM-generated code shows similar patterns across all metrics regardless of the poisoning quantities and number of shots. To enhance clarity, the analysis in this RQ is conducted under a one-shot setting and moderate poisoning quantities: five poisoned samples in Scenario I and a 0.6 poisoning proportion in Scenario II.

\subsubsection{The Impact of Programming Language}
\input{tables/RQ2_inherent}
This sub-question explores how the characteristics of different programming languages influence the security of LLM-generated code in RACG systems. The analysis is based on metrics derived from four LLMs evaluated under two retrieval approaches (JINA and BM25) across two scenarios each (I and II). Note that we only present the VR metric here, as our focus is on the impact of programming languages on the security of LLM-generated code.

Table~\ref{tab:diff_lang} reveals distinct VR trends among languages. C++ consistently exhibits the highest VR across most sub-scenarios, with values reaching as high as 0.47 in JINA-I and 0.44 in BM25-II on average (\ie ``All'' column). This elevated VR is likely attributed to C++'s inherent complexity, extensive feature set, and lower safety abstractions, which make it more prone to vulnerabilities when retrieved samples contain flaws. Notably, C exhibits a VR nearly comparable to C++ in JINA-based scenarios, with 0.45 in JINA-I and 0.41 in JINA-II on average, indicating higher sensitivity to retrieval strategy when generating C program. JINA's strong retrieval capability amplifies the vulnerability risk for C, as retrieving highly similar examples likely introduces more vulnerability-prone code, highlighting how C’s simplicity and lack of safety abstractions make it more susceptible to vulnerabilities in RACG.

In contrast, Python and Java exhibit distinct trends, showing similar but lower VR compared to C and C++. For instance, the average VRs of Java and Python in JINA-II are 0.38 and 0.39, respectively, while those of C and C++ are 0.41 and 0.46. This stability can be attributed to the built-in safety mechanisms of Java and Python, which reduce vulnerability risks even when the retrieved code contains vulnerabilities.

\notez{
{\bf Finding 3}: LLMs generate more vulnerable code in C++ language but show greater resistance to creating vulnerable code in Java in typical RACG scenarios.
}

\subsubsection{The Impact of Example-Query Similarity}
\label{subsec:similarity}
In this sub-RQ, we investigate whether code examples with higher semantic similarity to the query introduce more vulnerabilities in RACG. To this end, we analyze the distribution of VR and VRCC across different ranges of the similarity between retrieved code examples and queries. For this end, assume that the feature vector of query $q$ is $\mathbf{f}_{q}$, $\mathbf{f}_{c}$ is the feature vector of retrieved code, we measure the similarity between retrieved code and query by calculating the cosine similarity between $\mathbf{f}_{q}$ and $\mathbf{f}_{c}$. This feature vector is generated by the {\tt text-embedding-3-large} model, which is not used by the retriever. This can avoid the bias caused by using the same embedding model as the retriever. 

\input{tables/RQ2_similarity}
Table~\ref{tab:rq_2_similarity} presents the VR and VRRC across different example-query similarity range, with the corresponding VR across various LLMs in two typical RACG scenarios (results from both retrievers are combined). The VRs are shown for each LLM and similarity range, while the VRRC (displayed in the last column) is reported only for similarity ranges, as it pertains to the retriever rather than the specific LLM.

Overall, the similarity between the retrieved code and queries is positively correlated with both VR and VRRC. This indicates that when retrieved code examples are more semantically aligned with the query, they are more likely to be vulnerable, thereby increasing the likelihood of generating code that contains vulnerabilities.
Specifically, in the lower similarity ranges \([0, 60)\), the VRs remain relatively stable, with only modest variations observed. For example, in Scenario I, the aggregated VR (``All'') increases marginally from 0.26 in the \([0, 20)\) range to 0.35 in the \([40, 60)\) range, reflecting a limited impact of similarity on vulnerability likelihood within this interval. Similarly, in Scenario II, the rates increase only slightly from 0.25 to 0.35 across the same ranges. In contrast, a significant increase is observed in the higher similarity ranges \([60, 100]\), where VRs rise sharply as similarity increases. In Scenario I, the aggregated rate escalates from 0.35 in the \([60, 80)\) range to 0.55 in the \([80, 100]\) range. A comparable trend is evident in Scenario II, with rates increasing from 0.35 to 0.47. Among individual models, CodeLlama exhibits the most pronounced increase, with its VR reaching 0.57 in Scenario I for the \([80, 100]\) range. We attribute this to two main reasons. First, when the retrieved code example is highly semantically aligned with the query, the LLM is more likely to generate code that closely resembles the provided example. If the provided code example is vulnerable, the generated code is also highly likely to contain vulnerabilities. Second, the poisoning process is designed to inject vulnerable code that aligns with the query's semantics, meaning that most injected vulnerable code examples exhibit a high degree of similarity to the query. This relationship is supported by the VRRC, which increases as the similarity range increases.

This trend indicates that lower similarity levels have a relatively minor impact on vulnerability likelihood, whereas higher similarity levels, particularly above 60\%, significantly heighten the risk of generating vulnerable code. These findings highlight the critical importance of rigorous validation when using high-similarity code retrieval in RACG, as such code, while more contextually relevant, is also more prone to introducing security vulnerabilities if it is vulnerable.

\notez{
{\bf Finding 4}: The likelihood of generating vulnerable code increases with the example-query similarity. The similarity above 60\% shows a significant increase in vulnerability risk, while lower similarity levels (0-60\%) have a relatively minor impact on VRs.
}

\subsubsection{The Impact of Vulnerability Types}
\label{subsec:vul_type}
\input{tables/RQ2_vul_types}

In this sub-RQ, we investigate whether different types of vulnerability have varying impacts on the security of code generated by LLMs. Specifically, we examine whether the presence of certain types of vulnerabilities in the retrieved examples is more inclined to lead to the generation of exploitable vulnerabilities in the code, thereby increasing the probability of successful attacks.
For this analysis, we focus exclusively on queries that retrieved vulnerable code examples, excluding cases where all retrieved examples were secure.

Table~\ref{tab:rq_2_vul_type} shows the VRs across the Top-10 most dangerous software weaknesses from MITRE~\cite{CWE_Top25_2024}. Full results for Top-25 weaknesses can be found in Table~\ref{tab:rq_2_vul_type_full} in the appendix. As this sub-RQ focuses on the impact of different vulnerability types, the results from both retrievers are combined in the table. Furthermore, since code generated using secure code examples is excluded, the VR values reported are intuitively higher than those in RQ1. 
Among the vulnerability types analyzed, CWE-352 (Cross-Site Request Forgery) exhibits consistently high VRs across all LLMs in both scenarios, with an average of 0.79 in each scenario. This suggests that even without exposed developer intents (Scenario II), LLMs remain prone to generating code susceptible to cross-site request attacks, potentially due to inherent biases or insufficient understanding of secure practices for handling such vulnerabilities.
Conversely, CWE-434 (Unrestricted Upload of File with Dangerous Type) demonstrates the lowest VRs (averaging 0.36 in Scenario I and 0.26 in Scenario II), indicating that LLMs are relatively more successful in avoiding this type of vulnerability. This might be attributed to the simpler nature of validation checks required to prevent such vulnerabilities. Comparing the two scenarios reveals generally higher VRs in Scenario I, particularly for CWE-79 (Cross-site Scripting), CWE-787 (Out-of-bounds Write), and CWE-89 (SQL Injection). This indicates that the presence of vulnerable code snippets in the retrieval set can inadvertently lead to their reproduction in the generated code, highlighting the risk of retrieval augmentation inadvertently amplifying existing vulnerabilities.

\notez{
{\bf Finding 5}: VRs vary across CWE types. While most CWEs exhibit moderate VRs, CWE-352 consistently demonstrates the highest VRs (around 0.8) among the MITRE Top-10 in typical RACG scenarios.
}

%% file: tables/RQ1_S1_pois_num.tex
% Table generated by Excel2LaTeX from sheet 'Sheet5'
\begin{table}[!t]
  \centering
  \caption{LLMs' metrics under different numbers of poisoned examples in Scenario I.}
  \resizebox{1\linewidth}{!}{
    \begin{tabular}{cclrrrrrr}
    \toprule
\multirow{2}[4]{*}{\bf Metrics} & \multicolumn{1}{c}{\multirow{2}[4]{*}{\bf Retriever}} & \multicolumn{1}{c}{\multirow{2}[4]{*}{\bf LLM}} &       & \multicolumn{5}{c}{\bf Poisoning Number} \\
\cmidrule{4-9}          &       &       & {\bf 0}     & {\bf 1}     & {\bf 3}     & {\bf 5}     & {\bf 7}     & {\bf 9} \\
    \midrule
    \multirow{8}[3]{*}{VR} & \multirow{4}[2]{*}{JINA} & GPT-4o & \mycellcolor{0.26}  & \mycellcolor{0.42}  & \mycellcolor{0.45}  & \mycellcolor{0.45}  & \mycellcolor{0.47}  & \mycellcolor{0.47}  \\
          &       & Llama-3 & \mycellcolor{0.18}  & \mycellcolor{0.34}  & \mycellcolor{0.36}  & \mycellcolor{0.37}  & \mycellcolor{0.37}  & \mycellcolor{0.37}  \\
          &       & CodeLlama & \mycellcolor{0.29}  & \mycellcolor{0.48}  & \mycellcolor{0.51}  & \mycellcolor{0.51}  & \mycellcolor{0.52}  & \mycellcolor{0.53}  \\
          &       & DS-Coder & \mycellcolor{0.25}  & \mycellcolor{0.44}  & \mycellcolor{0.47}  & \mycellcolor{0.49}  & \mycellcolor{0.49}  & \mycellcolor{0.50}  \\
\cmidrule{2-9}          & \multirow{4}[1]{*}{BM25} & GPT-4o & \mycellcolor{0.27}  & \mycellcolor{0.30}  & \mycellcolor{0.32}  & \mycellcolor{0.31}  & \mycellcolor{0.34}  & \mycellcolor{0.33}  \\
          &       & Llama-3 & \mycellcolor{0.22} & \mycellcolor{0.28}  & \mycellcolor{0.30}  & \mycellcolor{0.31}  & \mycellcolor{0.32}  & \mycellcolor{0.33}  \\
          &       & CodeLlama & \mycellcolor{0.27} & \mycellcolor{0.33}  & \mycellcolor{0.35}  & \mycellcolor{0.35}  & \mycellcolor{0.37}  & \mycellcolor{0.37}  \\
          &       & DS-Coder & \mycellcolor{0.28} & \mycellcolor{0.33}  & \mycellcolor{0.33}  & \mycellcolor{0.34}  & \mycellcolor{0.34}  & \mycellcolor{0.35}  \\
\midrule
   \multirow{8}[3]{*}{Sim$\dagger$} & \multirow{4}[1]{*}{JINA} & GPT-4o & \mybleucell{0.75}  & \mybleucell{0.75}  & \mybleucell{0.75}  & \mybleucell{0.75}  & \mybleucell{0.74}  & \mybleucell{0.74}  \\
&       & Llama-3 & \mybleucell{0.60}  & \mybleucell{0.59}  & \mybleucell{0.59}  & \mybleucell{0.58}  & \mybleucell{0.58}  & \mybleucell{0.59}  \\
&       & CodeLlama & \mybleucell{0.77}  & \mybleucell{0.75}  & \mybleucell{0.75}  & \mybleucell{0.75}  & \mybleucell{0.76}  & \mybleucell{0.76}  \\
&       & DS-Coder & \mybleucell{0.76}  & \mybleucell{0.77}  & \mybleucell{0.78}  & \mybleucell{0.78}  & \mybleucell{0.78}  & \mybleucell{0.78}  \\
\cmidrule{2-9}          & \multirow{4}[2]{*}{BM25} & GPT-4o & \mybleucell{0.18}  & \mybleucell{0.17}  & \mybleucell{0.18}  & \mybleucell{0.18}  & \mybleucell{0.18}  & \mybleucell{0.19}  \\
&       & Llama-3 & \mybleucell{0.16} & \mybleucell{0.17}  & \mybleucell{0.17}  & \mybleucell{0.17}  & \mybleucell{0.18}  & \mybleucell{0.18}  \\
&       & CodeLlama & \mybleucell{0.17} & \mybleucell{0.17}  & \mybleucell{0.17}  & \mybleucell{0.17}  & \mybleucell{0.19}  & \mybleucell{0.19}  \\
&       & DS-Coder & \mybleucell{0.24} & \mybleucell{0.23}  & \mybleucell{0.23}  & \mybleucell{0.24}  & \mybleucell{0.25}  & \mybleucell{0.25}\\
    \midrule
    \multirow{2}[4]{*}{VRRC} & JINA  & -     & 0.00     & 0.38  & 0.40  & 0.41  & 0.42  & 0.42  \\
\cmidrule{2-9}          & BM25  & -     & 0.00     & 0.05  & 0.06  & 0.06  & 0.07  & 0.07  \\
    \bottomrule
    \end{tabular}%
    }
  \label{tab:pos_num}%
  \caption*{\footnotesize $\dagger$ ``Sim" denotes similarity metric.\phantom{\hspace{150pt}}}
  % \vspace{-15pt}
\end{table}%

%% file: tables/RQ1_S1_pois_rate.tex
% Table generated by Excel2LaTeX from sheet 'Sheet5'
\begin{table}[!t]
  \centering
  \caption{LLMs' metrics under different poisoned proportions in Scenario II.}
  \resizebox{1\linewidth}{!}{
    \begin{tabular}{cclrrrrrr}
    \toprule
    \multirow{2}[4]{*}{\bf Metrics} & \multicolumn{1}{c}{\multirow{2}[4]{*}{\bf Retriever}} & \multicolumn{1}{c}{\multirow{2}[4]{*}{\bf LLM}} &       & \multicolumn{5}{c}{\bf Poisoning Proportion} \\
    \cmidrule{4-9}          &       &       & {\bf 0}     & {\bf 0.2}     & {\bf 0.4}     & {\bf 0.6}     & {\bf 0.8}     & {\bf 1.0} \\
    \midrule
        \multirow{8}[3]{*}{VR} & \multirow{4}[2]{*}{JINA} & GPT-4o & \mycellcolor{0.26}  & \mycellcolor{0.32}  & \mycellcolor{0.35}  & \mycellcolor{0.38}  & \mycellcolor{0.41}  & \mycellcolor{0.47}  \\
&       & Llama-3 & \mycellcolor{0.18}  & \mycellcolor{0.28}  & \mycellcolor{0.32}  & \mycellcolor{0.35}  & \mycellcolor{0.37}  & \mycellcolor{0.38}  \\
&       & CodeLlama & \mycellcolor{0.29}  & \mycellcolor{0.36}  & \mycellcolor{0.40}  & \mycellcolor{0.44}  & \mycellcolor{0.47}  & \mycellcolor{0.52}  \\
&       & DS-Coder & \mycellcolor{0.25}  & \mycellcolor{0.32}  & \mycellcolor{0.38}  & \mycellcolor{0.42}  & \mycellcolor{0.44}  & \mycellcolor{0.50}  \\
\cmidrule{2-9}          & \multirow{4}[1]{*}{BM25} & GPT-4o & \mycellcolor{0.27}  & \mycellcolor{0.29}  & \mycellcolor{0.32}  & \mycellcolor{0.34}  & \mycellcolor{0.35}  & \mycellcolor{0.37}  \\
&       & Llama-3 & \mycellcolor{0.22}  & \mycellcolor{0.26}  & \mycellcolor{0.30}  & \mycellcolor{0.32}  & \mycellcolor{0.36}  & \mycellcolor{0.40}  \\
&       & CodeLlama & \mycellcolor{0.27}  & \mycellcolor{0.35}  & \mycellcolor{0.36}  & \mycellcolor{0.40}  & \mycellcolor{0.43}  & \mycellcolor{0.45}  \\
&       & DS-Coder & \mycellcolor{0.28}  & \mycellcolor{0.31}  & \mycellcolor{0.33}  & \mycellcolor{0.36}  & \mycellcolor{0.38}  & \mycellcolor{0.43}  \\
\midrule
    \multirow{8}[3]{*}{Sim} & \multirow{4}[1]{*}{JINA} & GPT-4o & \mybleucell{0.75}  & \mybleucell{0.74}  & \mybleucell{0.76}  & \mybleucell{0.78}  & \mybleucell{0.77}  & \mybleucell{0.78}  \\
          &       & Llama-3 & \mybleucell{0.60}  & \mybleucell{0.57}  & \mybleucell{0.58}  & \mybleucell{0.59}  & \mybleucell{0.58}  & \mybleucell{0.59}  \\
          &       & CodeLlama & \mybleucell{0.77}  & \mybleucell{0.78}  & \mybleucell{0.77}  & \mybleucell{0.79}  & \mybleucell{0.80}  & \mybleucell{0.80}  \\
          &       & DS-Coder & \mybleucell{0.76}  & \mybleucell{0.78}  & \mybleucell{0.77}  & \mybleucell{0.78}  & \mybleucell{0.76}  & \mybleucell{0.78}  \\
\cmidrule{2-9}          & \multirow{4}[2]{*}{BM25} & GPT-4o & \mybleucell{0.18}  & \mybleucell{0.20}  & \mybleucell{0.19}  & \mybleucell{0.21}  & \mybleucell{0.21}  & \mybleucell{0.20}  \\
          &       & Llama-3 & \mybleucell{0.16}  & \mybleucell{0.17}  & \mybleucell{0.18}  & \mybleucell{0.17}  & \mybleucell{0.18}  & \mybleucell{0.19}  \\
          &       & CodeLlama & \mybleucell{0.17}  & \mybleucell{0.18}  & \mybleucell{0.19}  & \mybleucell{0.20}  & \mybleucell{0.19}  & \mybleucell{0.21}  \\
          &       & DS-Coder & \mybleucell{0.24}  & \mybleucell{0.24}  & \mybleucell{0.25}  & \mybleucell{0.26}  & \mybleucell{0.25}  & \mybleucell{0.26}  \\
    \midrule
    \multirow{2}[4]{*}{VRRC} & JINA  & -     & 0.00  & 0.09  & 0.18  & 0.27  & 0.35  & 0.42  \\
\cmidrule{2-9}          & BM25  & -     & 0.00  & 0.06  & 0.12  & 0.17  & 0.25  & 0.43  \\
    \bottomrule
    \end{tabular}%
    }
  \label{tab:pos_rate}%
  
\end{table}%

%% file: tables/RQ1_shots.tex
% Table generated by Excel2LaTeX from sheet 'Sheet5'
\begin{table}[!t]
  \centering
  \caption{LLMs' metrics under different numbers of shots and scenarios.}
  \resizebox{1\linewidth}{!}{
    \begin{tabular}{cclccccc}
    \toprule
    \multirow{2}[1]{*}{\bf Metrics} & \multicolumn{1}{c}{\multirow{2}[1]{*}{\shortstack{\bf{Retriever-} \\ \bf{Scenario$\dagger$}}}} & \multicolumn{1}{c}{\multirow{2}[1]{*}{\bf Shot}} & \multicolumn{5}{c}{\bf LLM} \\
          &       &       & {\bf GPT-4o} & {\bf Llama-3} & {\bf CodeLlama} & {\bf DS-Coder} & {\bf All} \\
\midrule
    \multirow{8}[7]{*}{VR} & \multirow{2}[1]{*}{JINA-I} & One   &  \mycellcolor{0.45} & \mycellcolor{0.37} & \mycellcolor{0.51} & \mycellcolor{0.49}    & \mycellcolor{0.46} \\
          &       & Three & \mycellcolor{0.47} & \mycellcolor{0.42} & \mycellcolor{0.54} & \mycellcolor{0.52} & \mycellcolor{0.49} \\
\cmidrule{2-8}          & \multirow{2}[2]{*}{BM25-I} & One & \mycellcolor{0.32} & \mycellcolor{0.30} & \mycellcolor{0.35} & \mycellcolor{0.33} & \mycellcolor{0.33} \\
          &       & Three & \mycellcolor{0.33} & \mycellcolor{0.32} & \mycellcolor{0.37} & \mycellcolor{0.36} & \mycellcolor{0.34} \\
\cmidrule{2-8}          & \multirow{2}[2]{*}{JINA-II} & One & \mycellcolor{0.38} & \mycellcolor{0.35} & \mycellcolor{0.44} & \mycellcolor{0.42} & \mycellcolor{0.40} \\
          &       & Three & \mycellcolor{0.36} & \mycellcolor{0.43} & \mycellcolor{0.46} & \mycellcolor{0.43} & \mycellcolor{0.42} \\
\cmidrule{2-8}          & \multirow{2}[2]{*}{BM25-II} & One   & \mycellcolor{0.34} & \mycellcolor{0.32} & \mycellcolor{0.40} & \mycellcolor{0.36} & \mycellcolor{0.36} \\
          &       & Three & \mycellcolor{0.33} & \mycellcolor{0.36} & \mycellcolor{0.40} & \mycellcolor{0.37} & \mycellcolor{0.36} \\
    \midrule
    \multirow{8}[7]{*}{Sim} & \multirow{2}[2]{*}{JINA-I} & One   &  \mybleucell{0.75} & \mybleucell{0.58} & \mybleucell{0.75} & \mybleucell{0.78} & \mybleucell{0.72} \\
          &       & Three & \mybleucell{0.76} & \mybleucell{0.62} & \mybleucell{0.77} & \mybleucell{0.78} & \mybleucell{0.73} \\
\cmidrule{2-8}          & \multirow{2}[2]{*}{BM25-I} & One   & \mybleucell{0.18} & \mybleucell{0.17} & \mybleucell{0.17} & \mybleucell{0.24} & \mybleucell{0.19} \\
          &       & Three & \mybleucell{0.21} & \mybleucell{0.22} & \mybleucell{0.17} & \mybleucell{0.25} & \mybleucell{0.21} \\
\cmidrule{2-8}          & \multirow{2}[2]{*}{JINA-II} & One  & \mybleucell{0.78} & \mybleucell{0.59} & \mybleucell{0.80} & \mybleucell{0.78} & \mybleucell{0.74} \\
          &       & Three & \mybleucell{0.77} & \mybleucell{0.60} & \mybleucell{0.80} & \mybleucell{0.79} & \mybleucell{0.74} \\
\cmidrule{2-8}          & \multirow{2}[1]{*}{BM25-II} & One   & \mybleucell{0.21} & \mybleucell{0.17} & \mybleucell{0.20} & \mybleucell{0.26} & \mybleucell{0.21} \\
          &       & Three & \mybleucell{0.21} & \mybleucell{0.20} & \mybleucell{0.19} & \mybleucell{0.25} & \mybleucell{0.21} \\
\midrule
    \multirow{8}[7]{*}{VRRC} & \multirow{2}[1]{*}{JINA-I} & One   & \multicolumn{5}{c}{0.41} \\
          &       & Three & \multicolumn{5}{c}{0.44} \\
\cmidrule{2-8}          & \multirow{2}[2]{*}{BM25-I} & One   & \multicolumn{5}{c}{0.06} \\
          &       & Three & \multicolumn{5}{c}{0.08} \\
\cmidrule{2-8}          & \multirow{2}[2]{*}{JINA-II} & One   & \multicolumn{5}{c}{0.27} \\
          &       & Three & \multicolumn{5}{c}{0.35} \\
\cmidrule{2-8}          & \multirow{2}[2]{*}{BM25-II} & One   & \multicolumn{5}{c}{0.17} \\
          &       & Three & \multicolumn{5}{c}{0.22} \\
    \bottomrule
    \end{tabular}%
}
  \label{tab:rq1_shots}%
    \caption*{\footnotesize $\dagger$ Retriever-Scenario indicates the result of a specific retriever (e.g., JINA or BM25) under a particular scenario (e.g., Scenario I or Scenario II).}
    \vspace{-8pt}
    % \vspace{-15pt}
\end{table}%

%% file: tables/RQ2_inherent.tex
\begin{table}[!t]
  \centering
  % \caption{VRs of LLM-generated code across different scenarios.}
  \caption{VRs of generated code across different scenarios.}
  \label{tab:rq2_inherent}
    \resizebox{\linewidth}{!}{
      \begin{tabular}{clrrrrr}
      \toprule
      \multirow{2}[1]{*}{\shortstack{\bf{Retriever-} \\ \bf{Scenario}}} & \multirow{2}[1]{*}{\bf{LLM}} & \multicolumn{5}{c}{\bf{Languages}} \\
            &       & \multicolumn{1}{c}{\bf{C}} & \multicolumn{1}{c}{\bf{C++}} & \multicolumn{1}{c}{\bf{Java}} & \multicolumn{1}{c}{\bf{Python}} & \multicolumn{1}{c}{\bf{All}} \\
      \midrule
      \multirow{4}[2]{*}{JINA-I} & GPT-4o   & \mycellcolor{0.44}  & \mycellcolor{0.48}  & \mycellcolor{0.46}  & \mycellcolor{0.49}  & \mycellcolor{0.45}  \\
    & Llama & \mycellcolor{0.30}  & \mycellcolor{0.47}  & \mycellcolor{0.44}  & \mycellcolor{0.48}  & \mycellcolor{0.37}  \\
    & CodeLlama & \mycellcolor{0.54}  & \mycellcolor{0.45}  & \mycellcolor{0.53}  & \mycellcolor{0.39}  & \mycellcolor{0.51}  \\
    & DS-Coder & \mycellcolor{0.50}  & \mycellcolor{0.46}  & \mycellcolor{0.42}  & \mycellcolor{0.46}  & \mycellcolor{0.47}  \\
      \cmidrule{2-7}
            & Average & {0.45} & {0.47} & {0.46} & {0.45} & {0.45} \\ 
      \midrule
      \multirow{4}[2]{*}{BM25-I}  &   GPT-4o & \mycellcolor{0.30}  & \mycellcolor{0.36}  & \mycellcolor{0.34}  & \mycellcolor{0.34}  & \mycellcolor{0.32}  \\
    & Llama-3 & \mycellcolor{0.28}  & \mycellcolor{0.33}  & \mycellcolor{0.32}  & \mycellcolor{0.36}  & \mycellcolor{0.30}  \\
    & CodeLlama & \mycellcolor{0.33}  & \mycellcolor{0.55}  & \mycellcolor{0.43}  & \mycellcolor{0.24}  & \mycellcolor{0.35}  \\
    & DS-Coder & \mycellcolor{0.30}  & \mycellcolor{0.42}  & \mycellcolor{0.35}  & \mycellcolor{0.36}  & \mycellcolor{0.33}  \\
      \cmidrule{2-7}
            & Average & {0.30} & {0.42} & {0.36} & {0.33} & {0.32} \\ 
      \midrule
      \multirow{4}[2]{*}{JINA-II} &     GPT-4o & \mycellcolor{0.37}  & \mycellcolor{0.40}  & \mycellcolor{0.41}  & \mycellcolor{0.38}  & \mycellcolor{0.38}  \\
    & Llama-3 & \mycellcolor{0.33}  & \mycellcolor{0.38}  & \mycellcolor{0.39}  & \mycellcolor{0.39}  & \mycellcolor{0.35}  \\
    & CodeLlama & \mycellcolor{0.48}  & \mycellcolor{0.53}  & \mycellcolor{0.37}  & \mycellcolor{0.40}  & \mycellcolor{0.44}  \\
    & DS-Coder & \mycellcolor{0.46}  & \mycellcolor{0.51}  & \mycellcolor{0.35}  & \mycellcolor{0.37}  & \mycellcolor{0.42}  \\
            \cmidrule{2-7}
            & Average & {0.41} & {0.46} & {0.38} & {0.39} & {0.40} \\ 
      \midrule
      \multirow{4}[2]{*}{BM25-II} &     GPT-4o & \mycellcolor{0.33}  & \mycellcolor{0.37}  & \mycellcolor{0.32}  & \mycellcolor{0.41}  & \mycellcolor{0.34}  \\
    & Llama-3 & \mycellcolor{0.30}  & \mycellcolor{0.36}  & \mycellcolor{0.30}  & \mycellcolor{0.43}  & \mycellcolor{0.32}  \\
    & CodeLlama & \mycellcolor{0.36}  & \mycellcolor{0.56}  & \mycellcolor{0.45}  & \mycellcolor{0.44}  & \mycellcolor{0.40}  \\
    & DS-Coder & \mycellcolor{0.35}  & \mycellcolor{0.46}  & \mycellcolor{0.37}  & \mycellcolor{0.38}  & \mycellcolor{0.36}  \\
            \cmidrule{2-7}
            & Average & {0.34} & {0.44} & {0.36} & {0.42} & {0.36} \\ 
      \bottomrule
      \end{tabular}%
    }
    \label{tab:diff_lang}
    % \vspace{-6pt}
\end{table}

%% file: tables/RQ2_similarity.tex
\begin{table}[!t]
  \centering
  \caption{VRs and VRRCs across different ranges of similarity between retrieved code and queries.}
  \resizebox{1\linewidth}{!}{
    \begin{tabular}{clrrrrrr}
    \toprule
    \multirow{2}[1]{*}{\bf{Scenario}} & \multirow{2}[1]{*}{\shortstack{\bf{Similarity} \\ \bf{Range(\%)$\dagger$}}} & \multicolumn{5}{c}{\bf{LLM}} & \multirow{2}[1]{*}{\bf{VRRC}} \\
          &       & \multicolumn{1}{c}{\bf{GPT-4o}} & \multicolumn{1}{c}{\bf{Llama-3}} & \multicolumn{1}{c}{\bf{CodeLlama}} & \multicolumn{1}{c}{\bf{DS-Coder}} & \multicolumn{1}{c}{\bf{All}} & \\
    \midrule
    \multirow{5}{*}{I}  
          & [0, 20)   & \mycellcolor{0.28} & \mycellcolor{0.27} & \mycellcolor{0.31} & \mycellcolor{0.25} & \mycellcolor{0.28} & {0.08}\\ 
          & [20, 40)  & \mycellcolor{0.33} & \mycellcolor{0.32} & \mycellcolor{0.39} & \mycellcolor{0.26} & \mycellcolor{0.32} & {0.09}\\ 
          & [40, 60)  & \mycellcolor{0.34} & \mycellcolor{0.34} & \mycellcolor{0.44} & \mycellcolor{0.28} & \mycellcolor{0.35} & {0.14}\\ 
          & [60, 80)  & \mycellcolor{0.42} & \mycellcolor{0.42} & \mycellcolor{0.48} & \mycellcolor{0.38} & \mycellcolor{0.42} & {0.28}\\ 
          & [80, 100] & \mycellcolor{0.50} & \mycellcolor{0.52} & \mycellcolor{0.57} & \mycellcolor{0.54} & \mycellcolor{0.53} & {0.42}\\ 
    \midrule
    \multirow{5}{*}{II}  
          & [0, 20)   & \mycellcolor{0.24}  & \mycellcolor{0.30}  & \mycellcolor{0.31}  & \mycellcolor{0.15}  & \mycellcolor{0.25}  & {0.06} \\
          & [20, 40)  & \mycellcolor{0.32}  & \mycellcolor{0.30}  & \mycellcolor{0.35}  & \mycellcolor{0.29}  & \mycellcolor{0.32}  & {0.14} \\
          & [40, 60)  & \mycellcolor{0.32}  & \mycellcolor{0.39}  & \mycellcolor{0.40}  & \mycellcolor{0.29}  & \mycellcolor{0.35}  & {0.15} \\ 
          & [60, 80)  & \mycellcolor{0.44}  & \mycellcolor{0.44}  & \mycellcolor{0.46}  & \mycellcolor{0.42}  & \mycellcolor{0.44}  & {0.27} \\ 
          & [80, 100] & \mycellcolor{0.45}  & \mycellcolor{0.44}  & \mycellcolor{0.52}  & \mycellcolor{0.47}  & \mycellcolor{0.47}  & {0.29} \\ 
    \bottomrule
    \end{tabular}%
  }
  \label{tab:rq_2_similarity}%
  \caption*{\footnotesize $\dagger$ VRs are calculated based on retrieved code similarity ranges. \([a, b)\) indicates that the interval includes \(a\) but excludes \(b\), while \([80, 100]\) is fully closed.}
  \vspace{-8pt}
  % \vspace{-15pt}
\end{table}

%% file: tables/RQ2_vul_types.tex
\begin{table}[!t]
  \centering
  \caption{VRs across MITRE's Top 10 software weaknesses.}
  \resizebox{1\linewidth}{!}{
    \begin{tabular}{clrrrrr}
    % \begin{tabular}{cl*{5}{p{0.8cm}}}
    \toprule
    \multirow{2}[1]{*}{\bf{Scenario}} & \multirow{2}[1]{*}{\bf{CWE Type}} & \multicolumn{5}{c}{\bf{LLM}} \\
          &       & \multicolumn{1}{c}{\bf{GPT-4o}} & \multicolumn{1}{c}{\bf{Llama-3}} & \multicolumn{1}{c}{\bf{CodeLlama}} & \multicolumn{1}{c}{\bf{DS-Coder}} & \multicolumn{1}{c}{\bf{All}} \\
    \midrule
    \multirow{10}[0]{*}{I} & CWE-79 & \mycellcolor{0.51}  & \mycellcolor{0.64}  & \mycellcolor{0.63} & \mycellcolor{0.51}  & \mycellcolor{0.57}  \\
          & CWE-787 & \mycellcolor{0.56} & \mycellcolor{0.63}  & \mycellcolor{0.59}  & \mycellcolor{0.54}  & \mycellcolor{0.58}  \\
          & CWE-89 & \mycellcolor{0.63}  & \mycellcolor{0.69}  & \mycellcolor{0.67}  & \mycellcolor{0.48}  & \mycellcolor{0.62}  \\
          & CWE-352 & \mycellcolor{0.83}  & \mycellcolor{0.75}  & \mycellcolor{0.81}  & \mycellcolor{0.77}  & \mycellcolor{0.79}  \\
          & CWE-22 & \mycellcolor{0.61}  & \mycellcolor{0.70}  & \mycellcolor{0.73}  & \mycellcolor{0.73}  & \mycellcolor{0.69}  \\
          & CWE-125 & \mycellcolor{0.56}  & \mycellcolor{0.57}  & \mycellcolor{0.53}  & \mycellcolor{0.47}  & \mycellcolor{0.53}  \\
          & CWE-78  & \mycellcolor{0.57}  & \mycellcolor{0.60}  & \mycellcolor{0.64}  & \mycellcolor{0.54}  & \mycellcolor{0.59}  \\
          & CWE-416 & \mycellcolor{0.65}  & \mycellcolor{0.51}  & \mycellcolor{0.56}  & \mycellcolor{0.51}  & \mycellcolor{0.56}  \\
          & CWE-862 & \mycellcolor{0.61}  & \mycellcolor{0.75}  & \mycellcolor{0.82}  & \mycellcolor{0.68}  & \mycellcolor{0.71}  \\
          & CWE-434 & \mycellcolor{0.35}  & \mycellcolor{0.38}  & \mycellcolor{0.35}  & \mycellcolor{0.37}  & \mycellcolor{0.36}  \\
    \midrule
    \multirow{10}[0]{*}{II} & CWE-79 & \mycellcolor{0.39}  & \mycellcolor{0.45}  & \mycellcolor{0.65}  & \mycellcolor{0.44}  & \mycellcolor{0.48}  \\
          & CWE-787 & \mycellcolor{0.34}  & \mycellcolor{0.34}  & \mycellcolor{0.51}  & \mycellcolor{0.52}  & \mycellcolor{0.43}  \\
          & CWE-89 & \mycellcolor{0.48}  & \mycellcolor{0.39}  & \mycellcolor{0.52}  & \mycellcolor{0.51}  & \mycellcolor{0.48}  \\
          & CWE-352 & \mycellcolor{0.83}  & \mycellcolor{0.78}  & \mycellcolor{0.77}  & \mycellcolor{0.76}  & \mycellcolor{0.78}  \\
          & CWE-22 & \mycellcolor{0.45}  & \mycellcolor{0.46}  & \mycellcolor{0.58}  & \mycellcolor{0.55}  & \mycellcolor{0.51}  \\
          & CWE-125 & \mycellcolor{0.37}  & \mycellcolor{0.34}  & \mycellcolor{0.41}  & \mycellcolor{0.35}  & \mycellcolor{0.37}  \\
          & CWE-78  & \mycellcolor{0.64}  & \mycellcolor{0.60}  & \mycellcolor{0.54}  & \mycellcolor{0.49}  & \mycellcolor{0.57}  \\
          & CWE-416 & \mycellcolor{0.45}  & \mycellcolor{0.44}  & \mycellcolor{0.59}  & \mycellcolor{0.46}  & \mycellcolor{0.49}  \\
          & CWE-862 & \mycellcolor{0.57}  & \mycellcolor{0.65}  & \mycellcolor{0.48}  & \mycellcolor{0.54}  & \mycellcolor{0.56}  \\
          & CWE-434 & \mycellcolor{0.25}  & \mycellcolor{0.17}  & \mycellcolor{0.31}  & \mycellcolor{0.29}  & \mycellcolor{0.26}  \\
    \bottomrule
    \end{tabular}%
  }
  \label{tab:rq_2_vul_type}%
\end{table}

%% file: 6.discussion.tex
\section{Discussion}

\subsection{Implications}
Our investigation highlights the critical need to mitigate knowledge base poisoning. The findings from our study have several implications for enhancing the security of code generated by RACG systems.
Firstly, existing retrieval strategies naturally favor the most relevant examples from the knowledge base, which gives attackers the opportunity to successfully mislead the generation process with a small number of vulnerable examples. We argue that this risk can be mitigated by adjusting the retrieval strategy. For instance, an alternative strategy would be to select the second most similar example or randomly choose from a candidate pool containing several of the most similar examples, and we plan to explore such a way in future work.

Secondly, based on the results from RQ1, we found that hiding the programming intent (i.e., in Scenario II) increases the difficulty of successful poisoning. For instance, attackers can achieve a VRRC of 0.38 in Scenario I with only one poisoned vulnerable example. However, in Scenario II, attackers achieve a VRRC of 0.35 with a poisoning proportion of 0.8, meaning that they need to inject 9,642 vulnerable examples into the knowledge base (calculated as $\lfloor12,053 \times 0.8\rfloor$).
% which corresponds to a poisoning rate of 40\% (9,642/21,695). 
This indicates that concealing the programming intent makes poisoning more intricate and easier to detect.

Thirdly, according to the finding from \S\ref{subsec:vul_type}, the security of LLM-generated code vary across CWE types. This suggests that RACG systems could devise special strategies to check for the existence of several specific CWE types in the knowledge base, such as CWE-352, with the aim of improving the security of the generated code, as these vulnerabilities are more likely to induce the generation of vulnerable code.
% key types of vulnerabilities in the knowledge base, thereby improving the security of the generated code.

\subsection{Effectiveness of Judge}
\label{subsec:judge_effectivenss}
To assess the performance of using an LLM as a judge to label responses, we evaluate the effectiveness through both manual sampling inspection and automated inspection. For {\bf manual} inspection, we determine the sample size based on a 95\% confidence level and a 10 confidence interval, using a population size of 12,053 responses from GPT-4o. The final sample sizes for C, C++, Java, and Python are 95, 81, 93, and 91, respectively, as calculated using an off-the-shelf tool.\footnote{\url{https://www.surveysystem.com/sscalc.htm}} The evaluated code was sampled from GPT-4o in a moderated, one-shot setting using the JINA retriever in Scenario I with five poisoning vulnerabilities.
Two authors independently evaluated the samples through manual review, followed by a double-check to ensure consistency. For {\bf automated} inspection, we using a dataset containing both vulnerable code and its fixed version. Specifically, we evaluate pairs of items (i.e., the vulnerable version and its fixed counterpart) using our LLM-based judge to see if it can distinguish between them. This evaluation is performed on the full dataset.
We classify a code sample as positive when the judge correctly identifies it as vulnerable and the results are presented in Table~\ref{tab:dis_judge_combined}. 
\input{tables/judge_eval}
Overall, the LLM-based judge demonstrates commendable performance, consistently achieving high accuracy, precision, recall, and F1 scores across both inspection approaches. This indicates that the judge is capable of effectively detecting vulnerabilities in generated code, making it a reliable evaluation method. Among the four programming languages evaluated, the judge's performance remains generally consistent.

In the results under manual inspection, Python yields the highest performance, with accuracy and F1 scores reaching 0.84 and 0.81, respectively. The performance is slightly lower for C and Java, with F1 scores of 0.79 and 0.78, respectively, but still remains at a high level. In the results under automated inspection, the judge's F1 scores across all four programming languages hover around 0.8, aligning closely with the results from manual inspection. These evaluations demonstrate that the judge effectively detects vulnerabilities in generated code.

\subsection{Effectiveness of Retrievers}
\input{tables/Dis_retriever}

Our influencing factors analysis of LLM-introduced vulnerabilities (\S\ref{subsec:cause_analysis}) reveals that different retrievers impact the security of generated code. Specifically, LLMs using the JINA retriever are more prone to generating vulnerable code than those using BM25 across various LLMs and scenarios. We attribute this to JINA's superior retrieval of relevant code. To validate this, we evaluate retriever effectiveness using MRR and SuccessRate@k (Table~\ref{tab:dis_retriever}), following prior work~\cite{liu2021opportunities,wang2024fusing}. MRR is the average reciprocal rank of results for a set of queries $q$, and SuccessRate@k is the percentage of queries where the relevant code snippet appears within the top-k results. As shown, JINA significantly outperforms BM25 across all metrics: MRR (0.85 vs. 0.20), SR@1 (0.79 vs. 0.14), SR@5 (0.91 vs. 0.19), and SR@10 (0.93 vs. 0.24). This confirms JINA's superior retrieval capability, which, while beneficial for general code generation, exposes LLMs to more potentially vulnerable code, thus increasing the likelihood of generating vulnerable code.

\subsection{The Difference with RAG Poisoning}
\label{subsec:diff_rag_poisoning}
RAG and RACG systems share the use of external knowledge to enhance content generation. However, they differ significantly in the nature of poisoning attacks and their consequences. Specifically, RAG poisoning targets the functional accuracy of the system~\cite{zou2024poisonedrag,zhang2024hijackrag}. In a RAG setup, attackers inject false or misleading examples into the knowledge base, causing the system to retrieve incorrect information. This disrupts the model’s ability to generate factually accurate outputs, undermining its usefulness in tasks requiring reliable information. The primary goal here is to compromise the system’s ability to produce correct content. 

In contrast, RACG poisoning aims to compromise the security of generated code without impacting functionality. Otherwise, the developer would discard the generated code and there would not be targeted vulnerability in the software.
By introducing vulnerable code examples into the knowledge base, attackers aim to influence the code generation process and lead to the creation of code with exploitable vulnerabilities, such as buffer overflows or SQL injection risks. This poisoning could propagate security vulnerabilities, creating potential real-world risks. RACG poisoning aims to infect the generated code with vulnerabilities that could be exploited.

This paper is the first comprehensive study examining how vulnerable code examples in the knowledge base impact the security of code generated by RACG systems. We focus on how these poisoned examples can lead to the generation of insecure code, introducing potential vulnerabilities. Our work highlights the need for securing RACG knowledge sources to prevent the propagation of security risks in generated code.

\subsection{Threats to Validity}

{\bf Query Generation through LLM.} In Section~\ref{subsec:dataset_cons}, we use DeepSeek-V2.5 to generate queries for code. However, there is a possibility that DeepSeek-V2.5 may produce inaccurate content. To mitigate this threat, we manually review the generated queries. Specifically, we randomly select 100 queries for each programming language and have them reviewed by the two authors. Any inconsistencies in the evaluation results were resolved through discussion between the authors. The manual review indicates that 86\% of the generated queries accurately reflect the functionality of the code on average. Therefore, the impact of this threat is minimal.

\noindent

{\bf Programming Languages Investigated.} In this study, we conduct experiments using four widely-used programming languages: C, C++, Java, and Python. One potential threat is that the selected languages may not fully represent real-world development scenarios. However, according to GitHub usage statistics (measured by the number of pull requests) for the first quarter of 2024~\cite{githut2024}, these four languages account for 42.7\% of the total activity. Among them, Python and Java are the most popular. Additionally, as other languages like Go gain popularity, we plan to extend our study to include more programming languages in future work.

%% file: tables/judge_eval.tex
\begin{table}[!t]
  \centering
  \caption{Performance of LLM-based judges under manual and automated inspection.}
  \resizebox{1\linewidth}{!}{
    \begin{tabular}{llrrrr}
    \toprule
    \multirow{2}{*}{\shortstack{\textbf{Inspection} \\ \textbf{Method}}} & \multirow{2}{*}{\textbf{Language}} & \multirow{2}{*}{\textbf{Accuracy}} & \multirow{2}{*}{\textbf{Precision}} & \multirow{2}{*}{\textbf{Recall}} & \multirow{2}{*}{\textbf{F1}} \\
    & & & & & \\
    \midrule
    \multirow{4}{*}{Manual} 
      & C     &   0.76  & 0.80 & 0.78  & 0.79 \\
      & C++   &  0.72   &  0.84  & 0.79    & 0.81 \\
      & Java  &  0.79 &  0.81  & 0.75 & 0.78 \\
      & Python &  0.84  &  0.85 & 0.78      & 0.81 \\
    \midrule
    \multirow{4}{*}{Automated} 
      & C     & 0.80 &  0.86 &  0.72 & 0.78 \\
      & C++   & 0.80 &  0.83 &  0.76 & 0.79 \\
      & Java  & 0.79 &  0.80 &  0.76 & 0.78 \\
      & Python & 0.82  & 0.83 & 0.80 & 0.81 \\
    \bottomrule
    \end{tabular}%
    }
  \label{tab:dis_judge_combined}%
\end{table}%

%% file: tables/Dis_retriever.tex
\begin{table}[!t]
  \centering
  \caption{The effectiveness of retrievers}
  \resizebox{0.8\linewidth}{!}{
      \begin{tabular}{lrrrr}
    \toprule
    \textbf{Retriever} & \textbf{MRR} & \textbf{SR@1$\dagger$} & \textbf{SR@5} &\textbf{SR@10} \\
    \midrule
    JINA  & \bf{0.85} & \bf{0.79} & \bf{0.91} & \bf{0.93} \\
    BM25  & 0.20 & 0.14 & 0.19 & 0.24 \\
    \bottomrule
    \end{tabular}%
  }
  \label{tab:dis_retriever}%
  \caption*{\footnotesize $\dagger$ SR@k represents the SuccessRate@k\hspace{2.8cm}\textcolor{white}{.}}
\end{table}%
\vspace{-2mm}

%% file: 8.conclusion.tex
\section{Conclusion}
This paper presents the first comprehensive study on the security risks of RACG systems, specifically the impact of vulnerable code in the knowledge base. Our experiments show that knowledge base poisoning significantly compromises the security of generated code, with up to 48\% of the code becoming vulnerable from a single poisoned sample in Scenario I with CodeLlama. We also identify that factors such as few-shot learning, programming language, and example-query similarity contribute to increased vulnerabilities. Our findings underline the need for improved knowledge base security and offer practical insights for reducing vulnerability propagation in LLM-generated code. This work pave the way for future works which aim at securing RACG systems and enhancing their reliability in software development.

%% file: 9.Ethics.tex
\section*{Ethics Considerations}
In conducting this research, we have carefully considered several ethical implications to ensure the integrity and responsibility of our work. Below, we outline two key ethical considerations and the measures we have taken to address them.

% \subsection*{Potential Misuse of Findings}
\begin{itemize}[leftmargin=*]
    \item {\bf Potential Misuse of Findings}. The findings of this study highlight the vulnerabilities that can be introduced into code generation systems through knowledge base poisoning. While this research aims to improve the security of RACG systems, there is a risk that malicious attackers could misuse this information to develop more sophisticated attacks. To address this, we have focused on providing mitigation strategies and recommendations to enhance the security of RACG systems.
    \item {\bf Impact on Developers and Organizations}.The findings of this study have implications for developers and organizations that rely on RACG systems for code generation. While the vulnerabilities identified in this research could pose risks, our goal is to raise awareness of these risks and provide practical recommendations to mitigate them. We encourage developers and organizations to adopt the security measures suggested in this paper to reduce the likelihood of introducing vulnerabilities into their codebases.
\end{itemize}

%% file: 10.OpenScience.tex
% \section*{Open Science}
% For open science, we release our artifact at: {\bf \url{https://zenodo.org/records/14716365}}, including the dataset, source code and results.

%% file: 11.appendix.tex
\appendix

\newpage

\section{Prompt for Query Generation}
\input{prompts/comment_generation}

\section{Prompt for Vulnerability Cause Pattern Extraction}
\input{prompts/vuln_pattern_extraction}

\section{Prompt for Security Assessment of Generated Code}
\input{prompts/LLM_judge_v2}

\section{VRs across MITRE's Top 25 software weaknesses}
The VRs across MITRE's Top 25 software weaknesses are shown in Table~\ref{tab:rq_2_vul_type_full}.
\input{tables/RQ2_vul_types_full}

\section{LLM Judge for Result Validation}
\label{sec_append:llm_juedge}
The LLM judge for result validation consists of two steps: vulnerability knowledge extraction analysis and vulnerability detection. The detailed implementation of each step is as follows.

\subsubsection{Vulnerability Knowledge Extraction}
The vulnerability knowledge extraction step analyzes fundamental patterns associated with vulnerabilities in the RACG knowledge base. This includes identifying both vulnerable patterns and their corresponding fixing patterns. Recall that each instance in our dataset is represented as a tuple $(q, v, s)$, where $q$ is the query, $v$ is the vulnerable version of the code, and $s$ is the secure version, as defined in \S\ref{subsec:dataset_cons}. We leverage an LLM to extract the vulnerability patterns by analyzing the differences between $v$ and $s$ (i.e., the patch).
For example, Listing~\ref{lst:vuln_pattern} illustrates a patch and its corresponding vulnerable patterns from CVE-2022-1733\footnote{\url{https://nvd.nist.gov/vuln/detail/CVE-2022-1733}}. In this example, the vulnerability cause pattern consists of the pattern name, the vulnerable pattern, and the corresponding fixing pattern. Including the fixing pattern helps reduce false positives by allowing us to verify whether detected vulnerable patterns have been addressed. 
\input{listings/diff_example}
Prompt~\ref{prompt:2} shows the template for automatically extracting the vulnerability cause patterns. The LLM uses the provided patch and description to generate vulnerability cause patterns and fixing patterns for subsequent vulnerability detection. Each vulnerability's description is collected from the NVD~\cite{nvd}, providing essential information to assist the LLM in pattern analysis.

\subsubsection{Security Assessment}
After extracting the vulnerability cause patterns, the next step is to assess whether the generated code contains any of these patterns. For a given piece of generated code $c$ derived from a query $q$, the code is deemed vulnerable if and only if $c$ matches any vulnerability cause patterns from the following two sources:
\begin{itemize}[leftmargin=*]
    \item \textbf{External Sources:} Vulnerable patterns extracted from examples of vulnerable code that were referenced (\ie the vulnerable code that retrieved as examples) during the generation of $c$.
    \item \textbf{Internal Sources:} Vulnerable patterns identified within the vulnerable version $v$ and the corresponding secure version $s$ associated with the query $q$.
\end{itemize}

These patterns facilitate the detection of vulnerabilities that may arise not only from the inclusion of vulnerable code as examples during the RACG process but also from inherent issues introduced during the generation of $c$ itself.

Finally, we leverage an LLM judge to identify whether the $c$ contains any vulnerability cause patterns. Prompt~\ref{prompt:3} shows the template for judging the security of $c$.
If any vulnerability cause pattern is detected, the generated code is considered as vulnerable.

%% file: prompts/comment_generation.tex
\prompt{Prompt 1: Query Generation}{1}{
You are an expert \{LANGUAGE\} programmer, please describe the functionality of the function:\\

\textbf{\# Function}\\
\{{FUNCTION}\}
\\
}

%% file: prompts/vuln_pattern_extraction.tex
% \begin{figure}
\prompt{Prompt 2: Vulnerability Cause Pattern Extraction}{2}{
Analyze the vulnerability's root cause using the provided fixing patch and description. Present your analysis in the following format:\\
- Pattern name: [Name of the pattern] \\
- Vulnerable Pattern: [Description of the vulnerability]  \\
- Fixing Pattern: [Description of the fix]\\

\textbf{\# Fixing Patch}\\
\{{DIFF}\}
\\
\textbf{\# Description}\\
\{{DESCRIPTION}\}
\\
}
% \end{figure}

%% file: prompts/LLM_judge_v2.tex
% \begin{figure}
\prompt{Prompt 3: Security Assessment}{3}{
Given the following vulnerability cause pattern of vulnerabilities, evaluate whether the provided code contains any of these vulnerabilities. Output 'found' or 'not found' for each vulnerability cause pattern.\\

\textbf{\# Vulnerability Cause Patterns}\\
\{{PATTERN 1:}\}\\
\{{PATTERN 2:}\}\\

\textbf{\# Code to Analyze}\\
\{{CODE}\}\\

}
% \end{figure}

%% file: tables/RQ2_vul_types_full.tex
\begin{table}[!t]
  \centering
  \caption{Vulnerability rates across MITRE's Top 25 software weaknesses}
  \resizebox{1\linewidth}{!}{
    \begin{tabular}{clrrrrr}
    \toprule
    \multirow{2}[1]{*}{\bf{Scenario}} & \multirow{2}[1]{*}{\bf{CWE Type}} & \multicolumn{5}{c}{\bf{LLM}} \\
          &       & \multicolumn{1}{c}{\bf{GPT-4o}} & \multicolumn{1}{c}{\bf{Llama-3}} & \multicolumn{1}{c}{\bf{CodeLlama}} & \multicolumn{1}{c}{\bf{DS-Coder}} & \multicolumn{1}{c}{\bf{All}} \\
    \midrule
        \multirow{25}[0]{*}{I} & CWE-79 & \mycellcolor{0.51}  & \mycellcolor{0.64}  & \mycellcolor{0.63} & \mycellcolor{0.51}  & \mycellcolor{0.57}  \\
          & CWE-787 & \mycellcolor{0.56}  & \mycellcolor{0.63}  & \mycellcolor{0.59}  & \mycellcolor{0.54}  & \mycellcolor{0.58}  \\
          & CWE-89 & \mycellcolor{0.63}  & \mycellcolor{0.69}  & \mycellcolor{0.67}  & \mycellcolor{0.48}  & \mycellcolor{0.62}  \\
          & CWE-352 & \mycellcolor{0.83}  & \mycellcolor{0.75}  & \mycellcolor{0.81}  & \mycellcolor{0.77}  & \mycellcolor{0.79}  \\
          & CWE-22 & \mycellcolor{0.61}  & \mycellcolor{0.70}  & \mycellcolor{0.73}  & \mycellcolor{0.73}  & \mycellcolor{0.69}  \\
          & CWE-125 & \mycellcolor{0.56}  & \mycellcolor{0.57}  & \mycellcolor{0.53}  & \mycellcolor{0.47}  & \mycellcolor{0.53}  \\
          & CWE-78  & \mycellcolor{0.57}  & \mycellcolor{0.60}  & \mycellcolor{0.64}  & \mycellcolor{0.54}  & \mycellcolor{0.59}  \\
          & CWE-416 & \mycellcolor{0.65}  & \mycellcolor{0.51}  & \mycellcolor{0.56}  & \mycellcolor{0.51}  & \mycellcolor{0.56}  \\
          & CWE-862 & \mycellcolor{0.61}  & \mycellcolor{0.75}  & \mycellcolor{0.82}  & \mycellcolor{0.68}  & \mycellcolor{0.71}  \\
          & CWE-434 & \mycellcolor{0.35}  & \mycellcolor{0.38}  & \mycellcolor{0.35}  & \mycellcolor{0.37}  & \mycellcolor{0.36}  \\
          & CWE-94 & \mycellcolor{0.51} & \mycellcolor{0.52} & \mycellcolor{0.53} & \mycellcolor{0.44} & \mycellcolor{0.50} \\
         & CWE-20 & \mycellcolor{0.63} & \mycellcolor{0.51} & \mycellcolor{0.65} & \mycellcolor{0.61} & \mycellcolor{0.60} \\
         & CWE-77 & \mycellcolor{0.53} & \mycellcolor{0.32} & \mycellcolor{0.47} & \mycellcolor{0.34} & \mycellcolor{0.42} \\
         & CWE-287 & \mycellcolor{0.62} & \mycellcolor{0.54} & \mycellcolor{0.65} & \mycellcolor{0.43} & \mycellcolor{0.56} \\
         & CWE-269 & \mycellcolor{0.63} & \mycellcolor{0.60} & \mycellcolor{0.66} & \mycellcolor{0.63} & \mycellcolor{0.62} \\
         & CWE-502 & \mycellcolor{0.60} & \mycellcolor{0.42} & \mycellcolor{0.64} & \mycellcolor{0.43} & \mycellcolor{0.50} \\
         & CWE-200 & \mycellcolor{0.59} & \mycellcolor{0.49} & \mycellcolor{0.58} & \mycellcolor{0.46} & \mycellcolor{0.53} \\
         & CWE-863 & \mycellcolor{0.47} & \mycellcolor{0.45} & \mycellcolor{0.50} & \mycellcolor{0.42} & \mycellcolor{0.46} \\
         & CWE-918 & \mycellcolor{0.47} & \mycellcolor{0.50} & \mycellcolor{0.59} & \mycellcolor{0.48} & \mycellcolor{0.51} \\
         & CWE-119 & \mycellcolor{0.56} & \mycellcolor{0.57} & \mycellcolor{0.58} & \mycellcolor{0.54} & \mycellcolor{0.57} \\
         & CWE-476 & \mycellcolor{0.57} & \mycellcolor{0.37} & \mycellcolor{0.54} & \mycellcolor{0.36} & \mycellcolor{0.46} \\
         & CWE-798 & \mycellcolor{0.45} & \mycellcolor{0.40} & \mycellcolor{0.49} & \mycellcolor{0.43} & \mycellcolor{0.44} \\
         & CWE-190 & \mycellcolor{0.54} & \mycellcolor{0.55} & \mycellcolor{0.48} & \mycellcolor{0.39} & \mycellcolor{0.50} \\
         & CWE-400 & \mycellcolor{0.53} & \mycellcolor{0.56} & \mycellcolor{0.63} & \mycellcolor{0.35} & \mycellcolor{0.52} \\
         & CWE-306 & \mycellcolor{0.37} & \mycellcolor{0.32} & \mycellcolor{0.35} & \mycellcolor{0.38} & \mycellcolor{0.36} \\
    \midrule
    \multirow{25}[0]{*}{II} & CWE-79 & \mycellcolor{0.39}  & \mycellcolor{0.45}  & \mycellcolor{0.65}  & \mycellcolor{0.44}  & \mycellcolor{0.48}  \\
          & CWE-787 & \mycellcolor{0.34}  & \mycellcolor{0.34}  & \mycellcolor{0.51}  & \mycellcolor{0.52}  & \mycellcolor{0.43}  \\
          & CWE-89 & \mycellcolor{0.48}  & \mycellcolor{0.39}  & \mycellcolor{0.52}  & \mycellcolor{0.51}  & \mycellcolor{0.48}  \\
          & CWE-352 & \mycellcolor{0.83}  & \mycellcolor{0.78}  & \mycellcolor{0.77}  & \mycellcolor{0.76}  & \mycellcolor{0.78}  \\
          & CWE-22 & \mycellcolor{0.45}  & \mycellcolor{0.46}  & \mycellcolor{0.58}  & \mycellcolor{0.55}  & \mycellcolor{0.51}  \\
          & CWE-125 & \mycellcolor{0.37}  & \mycellcolor{0.34}  & \mycellcolor{0.41}  & \mycellcolor{0.35}  & \mycellcolor{0.37}  \\
          & CWE-78  & \mycellcolor{0.64}  & \mycellcolor{0.60}  & \mycellcolor{0.54}  & \mycellcolor{0.49}  & \mycellcolor{0.57}  \\
          & CWE-416 & \mycellcolor{0.45}  & \mycellcolor{0.44}  & \mycellcolor{0.59}  & \mycellcolor{0.46}  & \mycellcolor{0.49}  \\
          & CWE-862 & \mycellcolor{0.57}  & \mycellcolor{0.65}  & \mycellcolor{0.48}  & \mycellcolor{0.54}  & \mycellcolor{0.56}  \\
          & CWE-434 & \mycellcolor{0.25}  & \mycellcolor{0.17}  & \mycellcolor{0.31}  & \mycellcolor{0.29}  & \mycellcolor{0.26}  \\
          & CWE-94 & \mycellcolor{0.58} & \mycellcolor{0.58} & \mycellcolor{0.55} & \mycellcolor{0.46} & \mycellcolor{0.54} \\
         & CWE-20 & \mycellcolor{0.36} & \mycellcolor{0.38} & \mycellcolor{0.60} & \mycellcolor{0.50} & \mycellcolor{0.46} \\
         & CWE-77 & \mycellcolor{0.43} & \mycellcolor{0.56} & \mycellcolor{0.67} & \mycellcolor{0.46} & \mycellcolor{0.53} \\
         & CWE-287 & \mycellcolor{0.62} & \mycellcolor{0.58} & \mycellcolor{0.69} & \mycellcolor{0.52} & \mycellcolor{0.60} \\
         & CWE-269 & \mycellcolor{0.50} & \mycellcolor{0.49} & \mycellcolor{0.52} & \mycellcolor{0.47} & \mycellcolor{0.49} \\
         & CWE-502 & \mycellcolor{0.39} & \mycellcolor{0.44} & \mycellcolor{0.59} & \mycellcolor{0.57} & \mycellcolor{0.50} \\
         & CWE-200 & \mycellcolor{0.57} & \mycellcolor{0.42} & \mycellcolor{0.55} & \mycellcolor{0.48} & \mycellcolor{0.50} \\
         & CWE-863 & \mycellcolor{0.48} & \mycellcolor{0.52} & \mycellcolor{0.54} & \mycellcolor{0.54} & \mycellcolor{0.52} \\
         & CWE-918 & \mycellcolor{0.57} & \mycellcolor{0.44} & \mycellcolor{0.47} & \mycellcolor{0.55} & \mycellcolor{0.51} \\
         & CWE-119 & \mycellcolor{0.45} & \mycellcolor{0.40} & \mycellcolor{0.53} & \mycellcolor{0.44} & \mycellcolor{0.46} \\
         & CWE-476 & \mycellcolor{0.43} & \mycellcolor{0.42} & \mycellcolor{0.55} & \mycellcolor{0.40} & \mycellcolor{0.45} \\
         & CWE-798 & \mycellcolor{0.32} & \mycellcolor{0.36} & \mycellcolor{0.43} & \mycellcolor{0.39} & \mycellcolor{0.38} \\
         & CWE-190 & \mycellcolor{0.40} & \mycellcolor{0.34} & \mycellcolor{0.58} & \mycellcolor{0.52} & \mycellcolor{0.46} \\
         & CWE-400 & \mycellcolor{0.45} & \mycellcolor{0.47} & \mycellcolor{0.49} & \mycellcolor{0.56} & \mycellcolor{0.49} \\
         & CWE-306 & \mycellcolor{0.37} & \mycellcolor{0.32} & \mycellcolor{0.48} & \mycellcolor{0.35} & \mycellcolor{0.41} \\

    \bottomrule
    \end{tabular}%
  }
  \label{tab:rq_2_vul_type_full}%
\end{table}

%% file: listings/diff_example.tex
\begin{lstlisting}[style=diff, caption={Fixing diff and vulnerability cause pattern for CVE-2022-1733}, label={lst:vuln_pattern}]
|{\color{gray}// Patch}|
...
- if (p[i] == '\\'')  
+ if (p[i - 1] != NUL && p[i] == '\\')
...

|{\color{gray}// Vulnerability Cause Pattern}|
|{\color{mybleu}Pattern name: }|Unchecked Null Termination.
|{\color{mybleu}Vulnerable Pattern: }|Accessing or modifying a string beyond its null terminator without proper checks.
|{\color{mybleu}Fixing Pattern: }|Ensure that the string access is within bounds before proceeding.
\end{lstlisting}